\documentclass[a4paper,11pt]{article}
\pdfoutput=1 
\usepackage{jheppub} 
\usepackage[T1]{fontenc} 
\usepackage{epsfig}
\usepackage{caption}
\usepackage{subcaption}
\usepackage{color,soul}
\usepackage{float}
\usepackage{url}
\usepackage{slashed}
\usepackage{accents}
\usepackage{lipsum}
\usepackage{mathtools}
\usepackage{physics}

\usepackage[normalem]{ulem}
\def\beq{\begin{equation}}
\def\eeq{\end{equation}}
\def\beqn{\begin{eqnarray}}
\def\eeqn{\end{eqnarray}}

\newcommand{\mdm}{\text{m}_\text{$\chi$}}

\title{Asymmetric dark matter from semi-annihilation}
\author[a]{Avirup Ghosh,}
\author[b]{Deep Ghosh}
\author[b]{and Satyanarayan Mukhopadhyay}
\affiliation[a]{Regional Centre for Accelerator-based Particle Physics, Harish-Chandra Research Institute, HBNI, Chhatnag Road, Jhunsi, Allahabad - 211 019, India}
\affiliation[b]{School of Physical Sciences, Indian Association for the Cultivation of Science, Kolkata 700032, India}

\emailAdd{avirupghosh@hri.res.in}
\emailAdd{tpdg@iacs.res.in}
\emailAdd{tpsnm@iacs.res.in}

\preprint{HRI-RECAPP-2020-005}

\abstract{We show that a general semi-annihilation scenario, in which a pair of dark matter (DM) particles annihilate to an anti-DM, and an unstable state that can mix with or decay to standard model states, can lead to particle anti-particle asymmetry in the DM sector. The present DM abundance, including the CP-violation in the DM sector and the resulting present asymmetry are determined entirely by a single semi-annihilation process at next-to-leading order. For large CP-violation in this process, we find that a nearly complete asymmetry can be obtained in the DM sector, with the observed DM density being dominated by the (anti-)DM particle. The presence of additional pair-annihilation processes can modify the ratio of DM and anti-DM number densities further, if the pair-annihilation is active subsequent to the decoupling of the semi-annihilation. For such a scenario, the required CP-violation for generating the same present asymmetry is generically much smaller, as compared to the scenario with only semi-annihilation present. We show that a minimal model with a complex scalar DM with cubic self-interactions can give rise to both semi- and pair-annihilations, with the required CP-violation generated at one-loop level. We also find that the upper bound on the DM mass from S-matrix unitarity in the purely asymmetric semi-annihilation scenario, with maximal CP-violation, is around 15 GeV, which is much stronger than in the WIMP and previously considered asymmetric DM cases, due to the required large non-zero chemical potential for such asymmetric DM.}

\begin{document} 
\maketitle
\flushbottom

\section{Introduction and Summary}
The  production mechanism for dark matter (DM) particles in the early Universe span a broad range of possibilities, ranging from processes in the thermal bath, to non-thermal mechanisms. If the DM states were in local kinetic and chemical equilibrium in the cosmic plasma at some epoch, its number-changing reactions would determine its final abundance observed today. Such number changing interactions can take place either entirely within the dark sector, or may involve the standard model (SM) particles as well. Here, we assume the existence of some conserved discrete or continuous global symmetry that can distinguish between the two sectors. 

The DM states can in general be either self-conjugate or have a distinct anti-particle. In the latter case, the number densities of DM particles and anti-particles can be different, if there is a conserved charge carried by the DM states which has a non-zero density in the Universe~\cite{Weinberg:2008zzc}. The generation of such an asymmetry requires DM number violating interactions, processes that violate charge conjugation ($C$) and charge conjugation-parity ($CP$), and departures from thermal equilibrium in the early Universe. Such Sakharov conditions~\cite{ Sakharov:1967dj} are known to be realized in different ways in baryogenesis mechanisms to produce matter-antimatter asymmetry in the SM sector~\cite{Yoshimura:1978ex, Ignatiev:1978uf, Kuzmin:1985mm, Affleck:1984fy, Weinberg:1979bt, Yoshimura:1979gy, Fukugita:1986hr}. In general, the asymmetries in the dark sector and visible sector may or may not be related, and in the latter case the asymmetry generation in the dark sector can be independently studied. A large number of mechanisms have been proposed for generating asymmetric DM, many of which connecting the asymmetries in the visible and dark sectors~\cite{Nussinov:1985xr, Petraki:2013wwa, Zurek:2013wia, Kaplan:2009ag, Kaplan:1991ah, Buckley:2010ui, Shelton:2010ta, Ibe:2011hq, Falkowski:2011xh, MarchRussell:2011fi, Bhattacherjee:2013jca, Fukuda:2014xqa}. 

\begin{figure}
\centering
\includegraphics[scale=0.75]{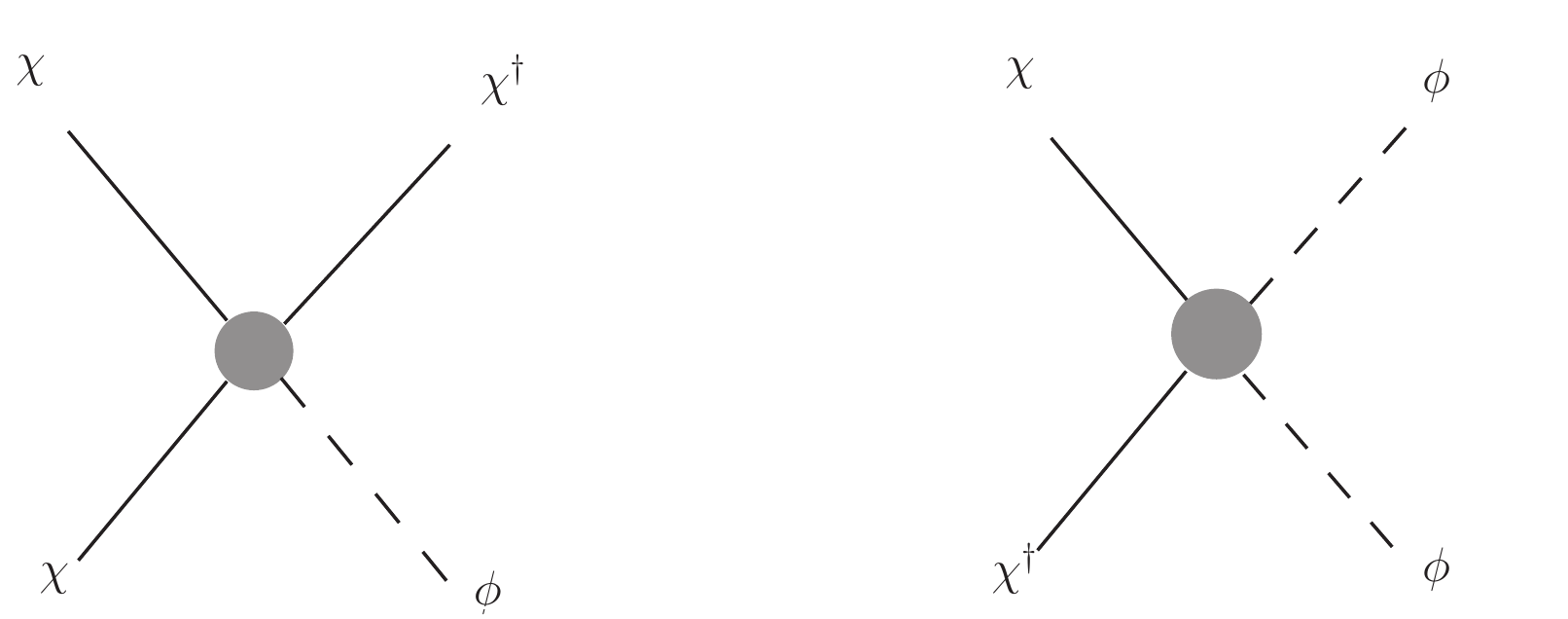}
\caption{\small{\em Effective interactions for the semi-annihilation (left) and pair-annihilation (right) processes. The former violates DM number by three units (and hence can generate a DM anti-DM asymmetry), while the latter conserves DM number.}}
\label{fig:eff_int}
\end{figure}

Among the DM number changing topologies, the simplest topologies with two DM, or two anti-DM, or one DM and one anti-DM particles in the initial state can involve either zero or one (anti-)DM particle in the final state, if there is a conserved stabilizing symmetry. The former final state corresponds to the standard pair-annihilation employed in the weakly interacting massive particle (WIMP) scenario, while the latter is the so-called semi-annihilation process~\cite{DEramo:2010keq}. If we assign a DM number of $n_\chi=1$ to the DM particle ($\chi$) and $n_\chi=-1$ to the anti-DM state ($\chi^\dagger$), then the annihilation of a $\chi \chi^\dagger$ pair does not change DM number $\Delta n_\chi = n_\chi^{\rm final} - n_\chi^{\rm initial} = 0$. On the other hand, a semi-annihilation process, for example, $\chi + \chi \rightarrow \chi^\dagger + \phi$, where $\phi$ is an unstable state not in the dark sector that can mix with or decay to SM states, can in general violate DM number (in the above reaction $\Delta n_\chi = -3$).  Thus, in the presence of semi-annihilations, the first Sakharov condition of DM number violation may easily be satisfied. We illustrate these effective interactions in Fig.~\ref{fig:eff_int}.

$CP-$violation in DM annihilation processes requires both the presence of residual complex phases in the Lagrangian (that cannot be removed by field re-definitions), as well as the interference between tree and loop level graphs, where the loop amplitudes develop a non-zero imaginary piece with intermediate states going on-shell. As we shall see in the subsequent discussion, the most minimal scenario with a complex scalar field dark matter with cubic self-interactions can satisfy both these requirements. This is one of the primary results of this paper. We compute the $CP-$violation that can be generated using this minimal setup, including the relevant loop-level amplitudes. 

The final Sakharov condition of out-of-equilibrium reactions can easily be realized in an expanding Universe, since the reaction time scales may become larger than the inverse Hubble scale at a given temperature, thereby leading to a departure from local thermal equilibrium. In our scenario, we achieve the out-of-equilibrium condition through the semi-annihilation process. As this process freezes out, a net difference in DM and anti-DM number densities is generated, starting from a symmetric initial condition. We formulate the set of coupled Boltzmann equations for the DM and anti-DM states, and study the evolution of their number densities as a function of the temperature scale to determine the resulting asymmetry, as well as the present net DM number density. 

As we shall see in the following, it is sufficient to have only the semi-annihilation process to generate a nearly maximal asymmetry in the DM sector with the required abundance, in which either only the DM or only the anti-DM survives in the present epoch. This is realized when the CP-violation in the process is large. For smaller CP-violation, the generated asymmetry is a partial one, with an unequal mixture of both DM and anti-DM states surviving. Thus in a scenario in which only the semi-annihilation process changes DM number in the thermal bath, or changes it sufficiently fast to achieve chemical equilibrium, this process entirely determines all the properties of asymmetric DM.

However, even in simple scenarios that realize the semi-annihilation process, including CP-violation through the interference of one-loop graphs with tree level ones, additional fast DM number-changing processes may also be present. In this class of models, there will be an interplay of semi-annihilation with these other processes in chemical equilibrium, such as the pair-annihilation process. In particular, if the semi-annihilation freezes out before the pair-annihilation, then the resulting ratio between DM and anti-DM co-moving number densities may be further enhanced. This results in the possibility that even with a tiny CP-violation in the DM sector, a maximal asymmetry may be achieved. Thus in this latter scenario one generically requires lower CP-violation for any amount of asymmetry, compared to the scenario in which only semi-annihilation is present.

Although studies on generating particle anti-particle asymmetries in both the matter sector and the dark matter sector have largely focussed on generating the asymmetries through CP-violating out-of-equilibrium decay of a particle (or multiple particles), asymmetry generation through CP-violating $2 \rightarrow 2$ annihilations has also been explored. This includes studies in baryogenesis and leptogenesis~\cite{Bento:2001rc,Nardi:2007jp,Gu:2009yx} and baryogenesis through WIMP annihilations~\cite{Cui:2011ab, Bernal:2012gv, Bernal:2013bga, Kumar:2013uca}, where the DM sector remains symmetric. In most previous studies on asymmetric DM, the primordial DM asymmetry is taken to be an input parameter, which is then evolved through the pair-annihilation process, using a set of coupled Boltzmann equations~\cite{Scherrer:1985zt, Griest:1986yu, Graesser:2011wi, Iminniyaz:2011yp, Lin:2011gj}.

The general possibility of generating particle anti-particle asymmetry in the dark sector from annihilations was studied in Refs.~\cite{Baldes:2014gca, Baldes:2015lka}. In particular, in Ref.~\cite{Baldes:2014gca} the general considerations of CPT and unitarity were imposed on a toy model involving two Dirac fermion fields in the dark sector pair-annihilating to the SM sector. In our study, however, we show that a minimal scenario with one complex scalar in the DM sector can lead to asymmetry generation through the semi-annihilation process. Furthermore, in Ref.~\cite{Baldes:2014gca}, the symmetric component of the DM was large at the end of asymmetry production, and it was necessary to introduce large particle antiparticle pair-annihilation cross-sections to remove this component. As discussed above, in our scenario, the pair-annihilation is not necessary to generate a DM asymmetry with the required abundance, but may be present in addition.

We now summarize the contents and the primary results of the subsequent sections. In Sec.~\ref{sec:mi}, we describe a model independent setup that encapsulates the role of the semi-annihilation process in generating a DM and anti-DM asymmetry in the present universe. We formulate a coupled set of Boltzmann equations involving the thermally averaged semi-annihilation rate, and a thermal average of the semi-annihilation rate times a suitably defined CP-violation parameter. {\em We find that for a large CP-violation, semi-annihilation alone gives rise to nearly complete asymmetry in the DM sector, with no symmetric component surviving at its decoupling}. For a given DM mass, larger the CP-violation, a correspondingly larger value of the semi-annihilation rate is required to satisfy the observed DM relic density. {\em Using S-matrix unitarity to bound the  semi-annihilation rate from above, we obtain an upper bound of $15$ GeV on the DM mass in this scenario, for maximal CP-violation and asymmetry.}

In Sec.~\ref{sec:interplay} we then introduce an additional number changing mechanism in the DM sector, namely the pair-annihilation process, and obtain the modified set of Boltzmann equations for this scenario to study the interplay of the two annihilation processes. We then go on to find a simple estimate of the present relic abundance in terms of the CP-violation, the annihilation rates and the dark matter mass. We obtain these estimates first in the case in which the symmetric component is completely annihilated away, and then compare it with results in which part of the symmetric component survives in the present Universe. {\em We find that in the presence of subsequent pair-annihilations, the required CP-violation to generate a complete DM asymmetry is much smaller, compared to the first scenario above with only semi-annihilation.} The required values of the pair-annihilation rates are also generically higher than in the standard WIMP scenario. {\em Imposing S-matrix unitarity to bound the pair-annihilation rate from above, we obtain an upper bound of around $25$ TeV on the DM mass, for a completely asymmetric scenario, which is to be contrasted with the result for only semi-annihilation above.} We show that a simple phase-diagram in the plane of the two annihilation rates summarizes the occurrence of symmetric and asymmetric DM, depending upon the values of these two rates. 

Finally, in Sec.~\ref{model} we describe a minimal example DM scenario that can lead to asymmetric DM production through the semi-annihilation mechanism, involving a complex scalar DM particle with a cubic self-interaction. The interplay of the semi- and pair-annihilation processes is realized in this scenario. We compute the CP-violation parameter explicitly in this model at one-loop level, and compare its values, and the correlation of the CP-violation parameter with the DM annihilation rates, with the ones obtained in the model-independent setup. We find that the required values of the physical parameters that can satisfy the observed DM abundance can be reproduced in this minimal scenario.

\section{Asymmetric dark matter from semi-annihilation}
\label{sec:mi}
To illustrate the main idea, we shall first consider the model independent parametrization of an example scenario involving only the semi-annihilation process, in which asymmetric dark matter through DM annihilations can be realized. The minimal number of DM degrees of freedom with which this can be implemented involves a complex scalar field ($\chi$). As mentioned in the Introduction, in the semi-annihilation process, two dark matter particles annihilate to produce an anti-dark matter particle and a neutral unstable state $\phi$: $\chi + \chi \rightarrow \chi^\dagger + \phi$. Here the state $\phi$ is not in the dark sector and can mix with or decay to standard model states. For production of on-shell $\phi$ particles from non-relativistic DM annihilation, we require $m_{\phi} < m_{\chi}$. We shall parametrize the next-to-leading-order cross-section for this process by $\sigma_S$, evaluated including the tree-level and one-loop diagrams. The corresponding CP-conjugate process is $\chi^\dagger + \chi^\dagger \rightarrow \chi +\phi$, with cross-section $\sigma_{\overline{S}}$, also evaluated at next-to-leading order. In general, since CP can be violated in the semi-annihilation process from the interference of the tree-level and one-loop graphs, $\sigma_{\overline{S}} \neq \sigma_S$. 

For temperatures $T>T_S$, where $T_S$ is the freeze-out temperature of the semi-annihilation process, using the conditions of detailed balance for the reactions $\chi + \chi \rightarrow \chi^\dagger + \phi$ and $\chi^\dagger + \chi^\dagger \rightarrow \chi +\phi$, we obtain the relation between the chemical potentials $\mu_\chi = \mu_{\chi^\dagger}=\mu_\phi$. For the cases when $\mu_\phi=0$, this implies that $\mu_\chi = \mu_{\chi^\dagger}=0$. During the freeze-out of the semi-annihilation, the third Sakharov condition of out-of-equilibrium is satisfied, and a DM anti-DM asymmetry may be generated. Since in this scenario for $T<T_S$, the DM particles are not in chemical equilibrium through any reactions, we do not assign it a chemical potential for these temperatures, but a pseudo-chemical potential may be defined as shown below in Eq.~\ref{eq:pseudo}. Furthermore, in this case, since no other number-changing processes are active for $T<T_S$, the  present particle anti-particle number density ratio ($n^0_\chi/n^0_{\chi^\dagger}$) is entirely determined by the semi-annihilation process. 

In addition to the cross-section $\sigma_S$, the other relevant parameters that determine the DM abundance are the mass of $\chi$ ($m_{\chi}$) and a CP-violation parameter $\epsilon$. Here, the CP-violation parameter is defined as:
\begin{equation}
\epsilon =\frac{ |M|^2_{\chi\chi\rightarrow \chi^{\dagger} \phi}-|M|^2_{\chi^{\dagger}\chi^{\dagger}\rightarrow \chi \phi}}{|M|^2_{\chi\chi\rightarrow \chi^{\dagger} \phi}+|M|^2_{\chi^{\dagger}\chi^{\dagger}\rightarrow \chi \phi}},
\label{eq:epsilon}
\end{equation}
where $|M|^2$ denotes the matrix element for the process. As for the cross-section difference between the CP-conjugate processes, the interference of the tree and one-loop amplitudes for the semi-annihilation process determines the value of $\epsilon$. 

The Boltzmann equation for the evolution of the DM number density $n_\chi$ can be expressed in terms of the squared matrix elements of the above processes as follows:
\begin{align}
\label{eq:boltz_semi}
\dfrac{dn_\chi}{dt}+3Hn_{\chi} &= -\int \prod^{4}_{i=1} \frac{d^3 p_i}{(2\pi)^3 2 E_{p_i}}g^2_{\chi}   (2 \pi)^4 \delta^{(4)}(p_1+p_2-p_3-p_4)
\bigg[2 f_{\chi}(p_1)f_{\chi}(p_2)\overline{|M|^2}_{\chi\chi\rightarrow \chi^{\dagger}\phi} \nonumber \\ 
&- 2 f_{\chi^{\dagger}}(p_3)f_{\phi}(p_4)\overline{|M|^2}_{\chi^{\dagger}\phi\rightarrow \chi\chi}
- f_{\chi^{\dagger}}(p_1)f_{\chi^{\dagger}}(p_2)\overline{|M|^2}_{\chi^{\dagger}\chi^{\dagger}\rightarrow \chi\phi}  \nonumber  \\ 
&+ f_{\chi}(p_3)f_{\phi}(p_4)\overline{|M|^2}_{\chi\phi\rightarrow \chi^{\dagger}\chi^{\dagger}} 
\bigg],
\end{align}
where $g_\chi$ denotes the number of internal degrees of freedom of $\chi$, and $\overline{|M|^2}$ is the squared matrix element for the given process, summed over final spins, and averaged over initial spins, {\em with appropriate factors for identical initial or final state particles included}. We can also write a similar Boltzmann equation for the evolution of the anti-particle number density $n_\chi^\dagger$, by replacing the symbol $\chi$ with the symbol $\chi^\dagger$ everywhere in Eqn.~\ref{eq:boltz_semi}. The distribution functions $f_i(p)$ in the above equation take the standard form 
\begin{equation}
f_{i}(p,t)=e^{-\frac{E_{i}}{T}}e^{\frac{\mu_{i}(t)}{T}},
\label{eq:pseudo}
\end{equation}
where we have set the Boltzmann constant $k_B=1$. The pseudo-chemical potential $\mu_{i}(t)$ parametrizes the small departure from the equilibrium distribution for the particle species $i$, and it approaches the chemical potential of the particle in chemical equilibrium~\cite{Dodelson:2003ft}. We note that CPT conservation can be used to relate the matrix elements for different processes above. For example, we have $|M|^2_{ \chi^{\dagger}\phi\rightarrow \chi\chi} =|M|^2_{ \chi^{\dagger}\chi^{\dagger}\rightarrow \chi\phi}$, where, since we are dealing with scalar particles only, the helicities of the states do not appear. 

Using energy conservation for the initial and final state particles,  and defining dimensionless variables (namely, $Y_i=n_i/s$ and  $x=m_\chi/T$, where $s$ is the entropy density per comoving volume), the coupled set of Boltzmann equations for the dark matter particle and anti-particle number densities take the following form:
\begin{eqnarray}
\dfrac{d Y_{\chi}}{d x} &=& -\dfrac{s}{H x}\left[A_S\left(Y^2_{\chi}+\dfrac{Y_0 Y_{\chi}}{2}\right)-B_S\left(\dfrac{Y^2_{\chi^{\dagger}}} {2}+Y_0 Y_{\chi^{\dagger}}\right)\right]  \nonumber  \\ 
\dfrac{d Y_{\chi^{\dagger}}}{d x} &=& -\dfrac{s}{H x}\left[B_S\left(Y^2_{\chi^{\dagger}}+\dfrac{Y_0 Y_{\chi^{\dagger}}}{2}\right)-A_S\left(\dfrac{Y^2_{\chi}} {2}+Y_0 Y_{\chi}\right)\right].
\label{boltz1_semi}
\end{eqnarray}  
Here, $H$ is the Hubble constant. We have also defined $A_S = \expval{\sigma v}_S+\expval{\epsilon \sigma v}_{S}$ and $B_S = \expval{\sigma v}_S-\expval{\epsilon \sigma v}_{S}$, with $\expval{\sigma v}_S$ and $ \expval{\epsilon \sigma v}_{S}$ being the thermally averaged cross-sections for the semi-annihilation process, without and with the asymmetry factor $\epsilon(p_i)$ included, respectively. In particular, 
\begin{equation}
\expval{\epsilon \sigma v}_{s} = \dfrac{\int \prod^{4}_{i=1} \frac{d^3 p_i}{(2\pi)^3 2 E_{p_i}}  (2 \pi)^4 \delta^{(4)}(p_1+p_2-p_3-p_4) \epsilon(p_i) \overline{|M_0|^2}f_0(p_1)f_0(p_2)}{\int \dfrac{d^3 p_1}{(2\pi)^3} \dfrac{d^3 p_2}{(2\pi)^3} f_0(p_1)f_0(p_2)} \hspace{0.5cm}
\end{equation}
with $f_0(p)=e^{-\frac{E}{T}}$ being the equilibrium distribution function when the chemical potential vanishes, and 
\begin{equation}
|M_0|^2 = |M|^2_{\chi\chi\rightarrow \chi^{\dagger} \phi}+|M|^2_{\chi^{\dagger}\chi^{\dagger}\rightarrow \chi \phi}.
\end{equation}
Finally, $Y_0$ is defined as $Y_0 =  \frac{1}{s} \int  \frac{d^3 p_i} {(2\pi)^3 } g_{\chi} f_0(p)$. We have assumed that throughout the evolution of the $\chi$ and $\chi^\dagger$ particles until the freeze-out of the semi-annihilation processes, the $\phi$ particle is in thermal equilibrium with the SM plasma with a vanishing chemical potential. We note that the equilibrium distribution with zero chemical potential $Y_0$ is not a solution of the coupled Eqs.~\ref{boltz1_semi}. This is because only the CP-violating process $\chi \chi \rightarrow \chi^\dagger \phi$ and its conjugate have been included while writing the collision term here. In other words, Eqs.~\ref{boltz1_semi} are valid when all the other processes in the thermal bath involving the $\chi$ and $\chi^\dagger$ particles have decoupled, by which time $Y_0$ is no longer a solution to the Boltzmann equations by the Boltzmann H-theorem~\cite{Weinberg:1995mt}. At even higher temperatures there must be other such processes with the same initial states, in order for the T-matrix element sum rules to be consistent with the requirements of CPT and S-matrix unitarity.

\subsection{Results}
In order to determine the DM relic abundance in a model-independent setup, we consider the thermally averaged cross-section for the semi-annihilation process ($\langle \sigma v \rangle_S$) as a free parameter. In addition, we define an effective CP-violation parameter $\epsilon_{\rm eff}=\langle \epsilon \sigma v \rangle_S/\langle \sigma v \rangle_S$. Therefore, there are three parameters appearing in the Boltzmann equations determining the DM and anti-DM number densities, as shown in Eq.~\ref{boltz1_semi}, namely, $m_\chi$, $\langle \sigma v \rangle_S$ and $\epsilon_{\rm eff}$. We see from Eq.~\ref{eq:epsilon} that $0<\epsilon<1$, whereby $\epsilon=0$ corresponds to no CP-violation in the semi-annihilation process, and $\epsilon=1$ to maximal CP-violation. We note that in general since $\epsilon$ is a function of the four-momenta of the particles, $\epsilon$ and $\epsilon_{\rm eff}$ are different. However, when the annihilation rates are dominated by the s-wave contributions, they become equal, and independent of the temperature. We shall work in this approximation in the model-independent analyses in Sec.~\ref{sec:mi} and Sec.~\ref{sec:interplay}.

\begin{figure}[t!]
\centering
\includegraphics[scale=0.55]{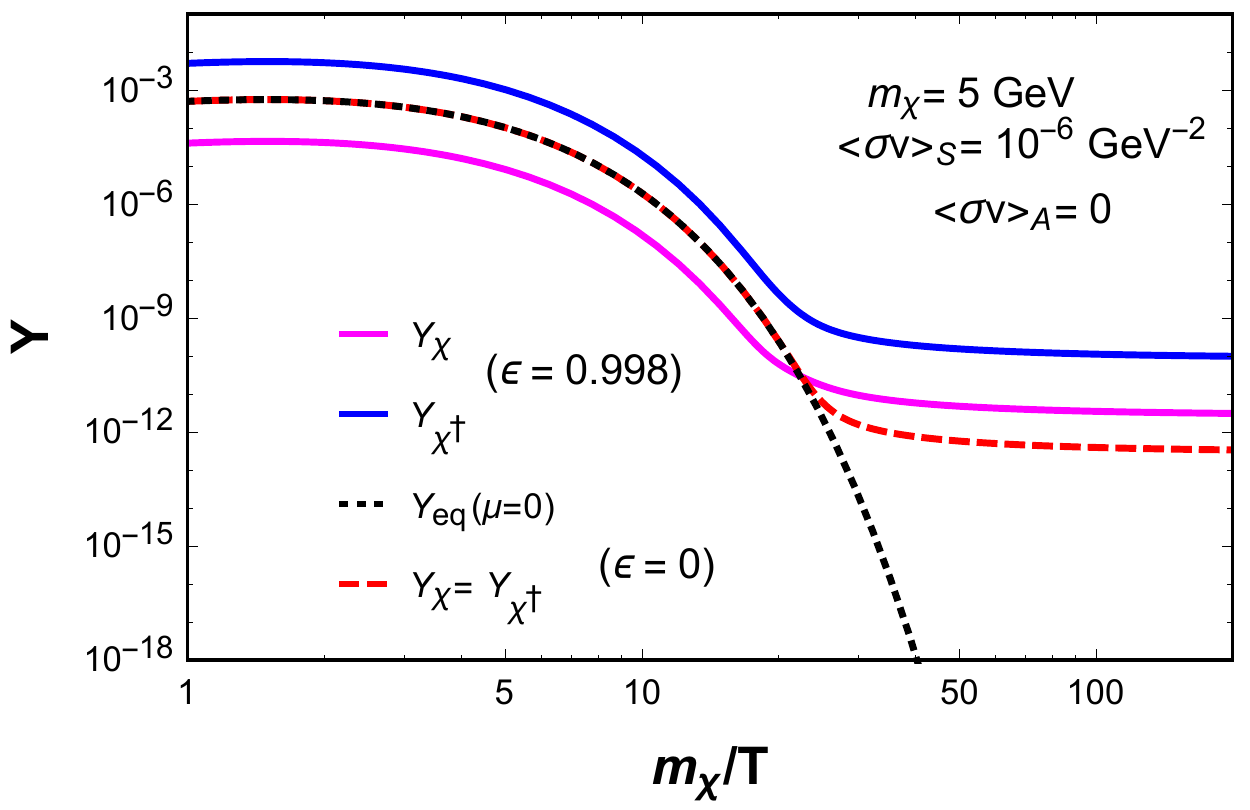} \hspace{0.9cm}
\includegraphics[scale=0.55]{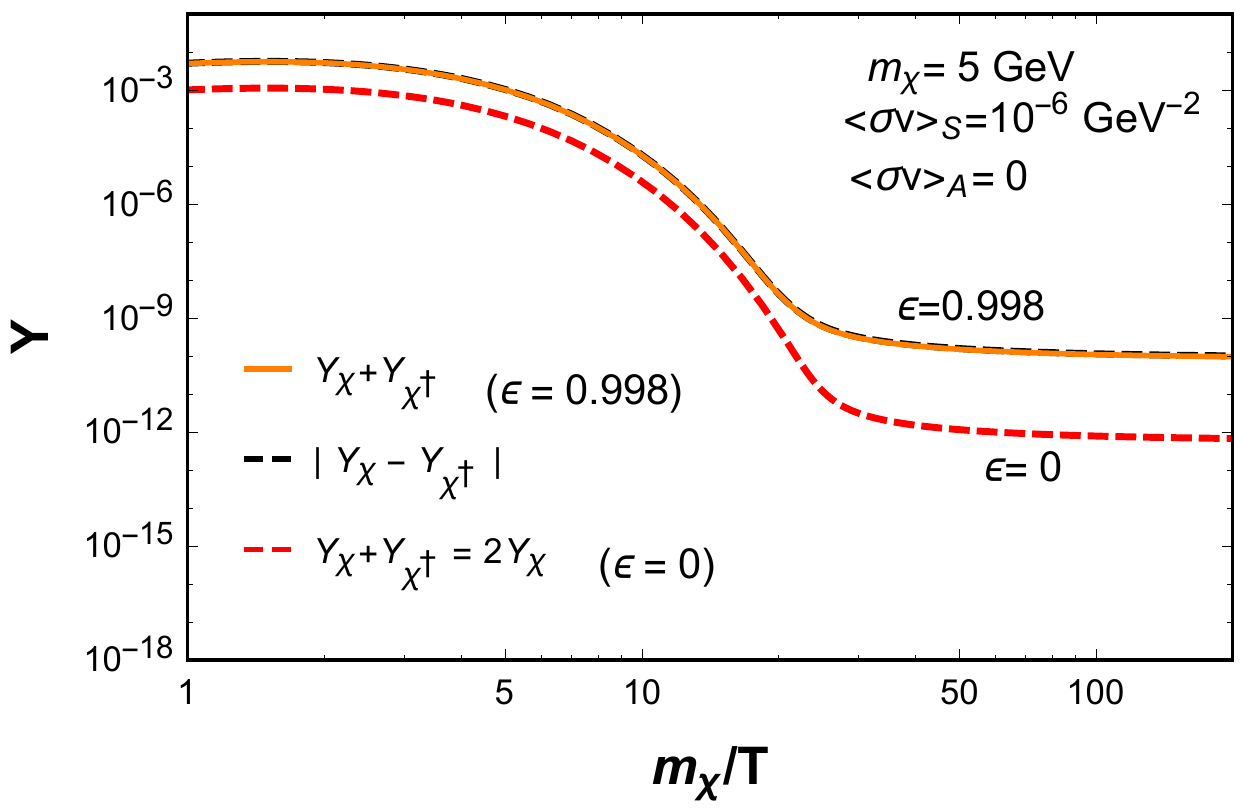}
\caption{\small{\em Dark matter ($Y_{\chi}$) and anti-dark matter yields ($Y_{\chi^\dagger}$) as a function of $x=m_\chi/T$, for the scenarios with no CP-violation ($\epsilon=0$) and nearly maximal CP-violation ($\epsilon=0.998$). The individual yields are shown in the left figure, while their sum and difference are shown in the right. See text for details.}}
\label{fig:semi_results0}
\end{figure}

We numerically solve the coupled Boltzmann equations in Eq.~\ref{boltz1_semi} to understand the parameter space in which the observed relic density of DM can be obtained. In Fig.~\ref{fig:semi_results0} (left panel), we show the dark matter ($Y_{\chi}$) and anti-dark matter yields ($Y_{\chi^\dagger}$) as a function of $x = m_\chi/T$, for the scenarios with no CP-violation ($\epsilon=0$, red dashed line) and nearly maximal CP-violation ($\epsilon=0.998$, pink and blue solid lines for DM and anti-DM respectively). Here, the semi-annihilation rate has been fixed to be $\langle \sigma v \rangle_S = 10^{-6} ~{\rm GeV}^{-2}$, with the DM mass $m_\chi=5$ GeV, to reproduce the observed central value of the DM relic density  $\Omega h^2 = 0.12$~\cite{Aghanim:2018eyx}. To contrast the results of this section with the ones in the next, in which we shall introduce DM pair-annihilation as a possible additional number-changing reaction, we have explicitly noted in this figure that the pair-annihilation rate vanishes in this scenario, i.e., $ \langle \sigma v \rangle_A=0$. In the left panel, we also show the corresponding equilibrium abundance $Y_{\rm eq}$ for the zero chemical potential scenario ($\mu=0$, black dashed line). 

In the right panel, we see that for $\epsilon=0.998$, the $\left(Y_{\chi}+Y_{\chi^\dagger}\right)$ (orange solid) and $|Y_{\chi}-Y_{\chi^\dagger}|$ (black dashed) lines almost identically trace each other. This demonstrates that for large $\mathcal{O}(1)$ CP-violation, the generated asymmetry is nearly maximal, and therefore the  (anti-)dark matter dominates the net yield. We also see from this figure that the total DM and anti-DM yield in the large CP-violation scenario ($\epsilon=0.998$) is larger than the yield for the CP-conserving case ($\epsilon=0$, red dashed line), for the same semi-annihilation rate $\langle \sigma v \rangle_S$. This is because of the large non-zero DM chemical potential in the CP-violating case. Therefore, in order to reproduce the observed relic abundance, we 
require correspondingly larger values of the semi-annihilation rate in the scenario with CP-violation.

\begin{figure}[t!]
\centering
\includegraphics[scale=0.5]{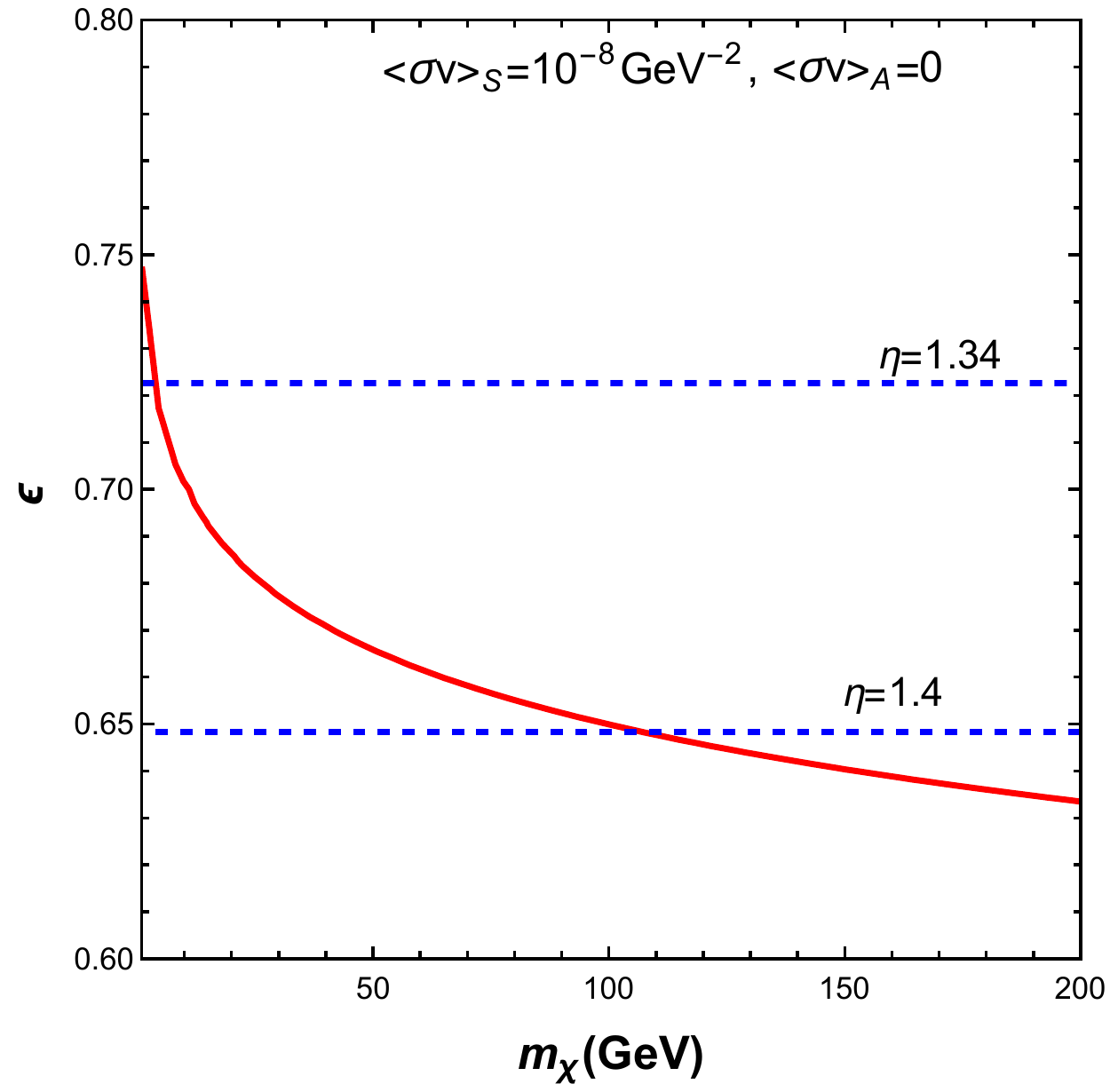}
\includegraphics[scale=0.48]{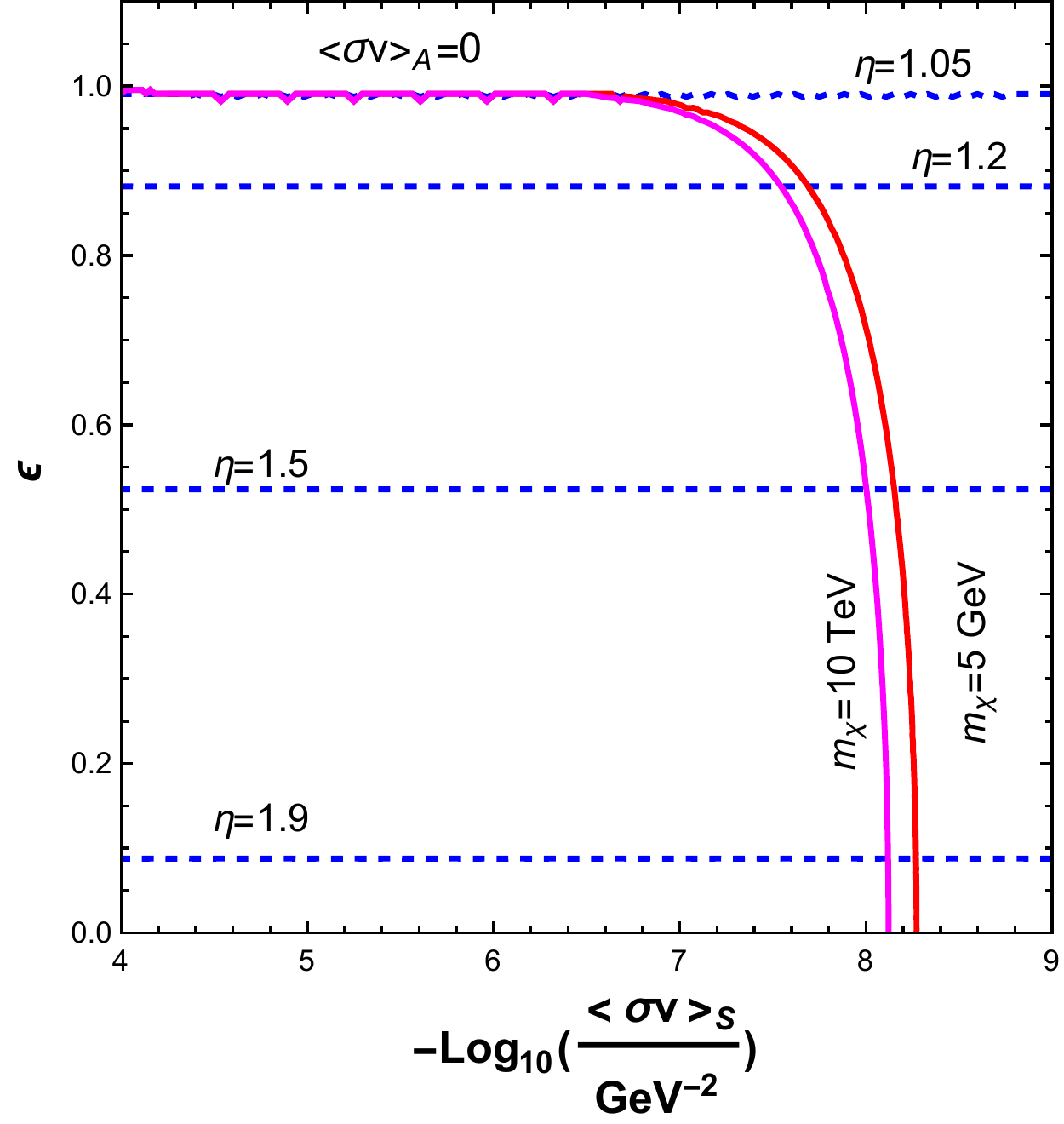}
\caption{\small{\em Contour in the $m_\chi - \epsilon$ plane (left panel, red solid line) in which the central value of the DM relic density  $\Omega h^2 = 0.12$ is reproduced, for a fixed value of the semi-annihilation rate. Same contours in the $\langle \sigma v \rangle_S - \epsilon$ plane for two different values of $m_\chi$ (right panel, red and pink solid lines). Also shown in both panels are contours of constant values of the relative abundance of DM and anti-DM, parametrized as $\eta = (Y_{\chi} (\infty)+Y_{\chi^\dagger}(\infty))/Y_{\chi^\dagger}(\infty)$, with $\eta \rightarrow 1$ being completely asymmetric DM. Only the semi-annihilation process has a non-zero rate in both the figures. See text for details.}}
\label{fig:semi_results1}
\end{figure}

In Fig.~\ref{fig:semi_results1} (left), we show the contour in the $m_\chi - \epsilon$ plane (red solid line) in which the central value of the DM relic density  $\Omega h^2 = 0.12$ is reproduced. For this figure, we have fixed the value of the semi-annihilation rate to be $\langle \sigma v \rangle_S = 10^{-8} ~{\rm GeV}^{-2}$. As before, for both the figures $ \langle \sigma v \rangle_A=0$. We also show contours in the $m_\chi - \epsilon$ parameter space for constant values of the relative abundance of DM and anti-DM, parametrized as
\begin{equation}
\eta = \frac{Y_{\chi} (\infty)+Y_{\chi^\dagger}(\infty)}{Y_{\chi^\dagger}(\infty)}, 
\label{eq:eta}
\end{equation}
where, the yield $Y_{\chi}(x)$ is evaluated at the present epoch with $x \rightarrow \infty$. Since for $\epsilon>0$, only  the $\chi^\dagger$ states survive for a scenario in which the symmetric component is completely annihilated away, in this limit, $\eta \rightarrow 1$. In scenarios in which the symmetric component partially survives, $1 < \eta <  2$. As we can see from this figure, for a fixed value of 
$\langle \sigma v \rangle_S$, higher values of $\epsilon$ imply a lower DM mass $m_\chi$ in which the relic density is reproduced. This is because, higher the CP-violation $\epsilon$, the higher is the difference in the number densities of the DM and anti-DM particles, which in turn implies a large pseudo-chemical potential. For a fixed value of the semi-annihilation rate, this also implies that the resulting frozen out number densities are higher, thus requiring a lower DM mass to saturate the same DM abundance. As is also clear, higher $\epsilon$ implies values of the relative abundance parameter $\eta$ closer to $1$. 

For a fixed DM mass, if we in turn keep increasing the CP violation $\epsilon$, the reaction rate $\langle \sigma v \rangle_S$ also needs to be correspondingly higher, for the same reason as described above. This is shown in Fig.~\ref{fig:semi_results1} (right), where for two fixed values of $m_\chi$ ($5$ GeV and $10$ TeV), we show the contours in the $\langle \sigma v \rangle_S - \epsilon$ plane (red and pink solid lines respectively) in which the central value of the DM relic density  $\Omega h^2 = 0.12$ is reproduced. The approach to $\epsilon=1$ in this figure is asymptotic, where the small numerical differences are not clear from the plot shown (which, however, we have checked numerically). For $\epsilon \rightarrow 1$, we see from this figure that $\eta \rightarrow 1$, with the surviving DM state being almost entirely the anti-DM.

\begin{figure}[t!]
\centering
\includegraphics[scale=0.5]{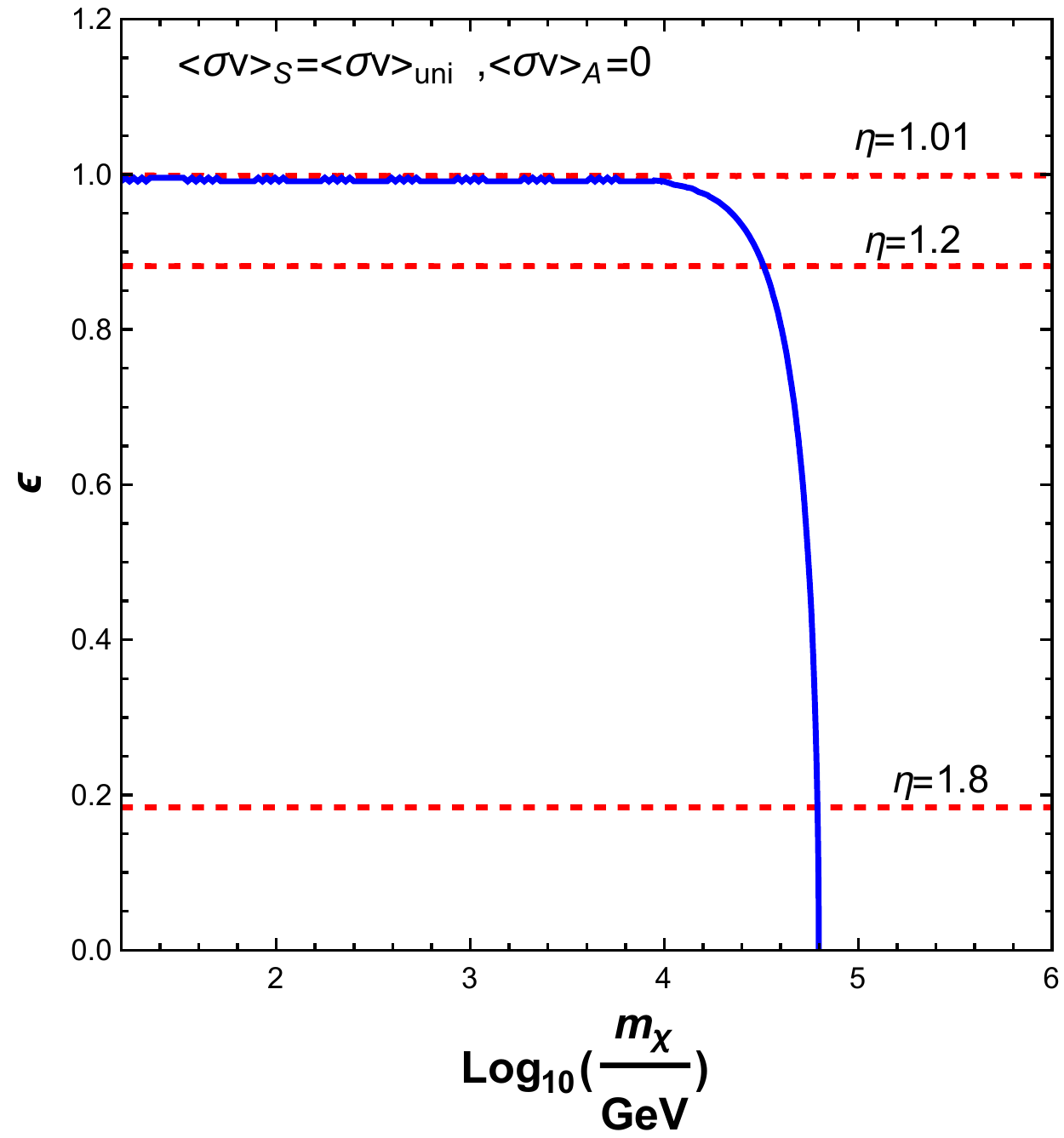}
\includegraphics[scale=0.55]{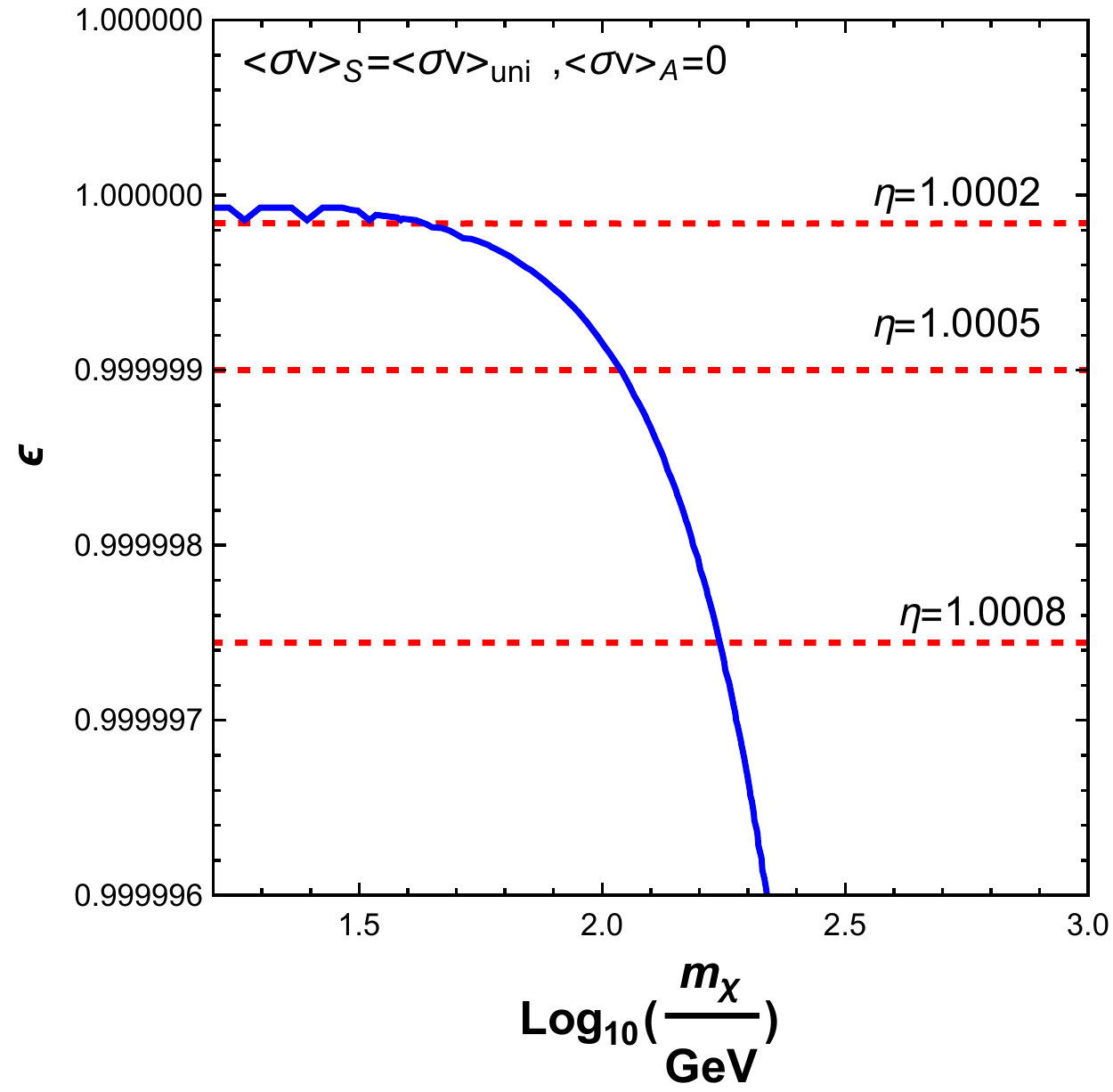}
\caption{\small{\em Contours in the $m_\chi - \epsilon$ plane in which the central value of the DM relic density  $\Omega h^2 = 0.12$ is reproduced, with the semi-annihilation rate fixed at its s-wave upper bound implied by S-matrix unitarity, $\langle \sigma v \rangle_S = \langle \sigma v \rangle_{\rm uni} = (4 \pi/ m_\chi^2)(x_F/ \pi)^{1/2}$. The $\eta$ parameter is the same as defined for Fig.~\ref{fig:semi_results1}, with $\eta \rightarrow 1$ being completely asymmetric DM. The right panel is the zoomed in version of the left panel focussing on a narrower region in the CP-violation parameter $\epsilon$. See text for further details.}}
\label{fig:semi_results2}
\end{figure}

How high can we go in the rate $\langle \sigma v \rangle_S$? We can use partial-wave S-matrix unitarity to bound the semi-annihilation cross-section from above. This in turn will also translate into an upper bound for possible values of the dark matter mass. The maximum allowed value of the cross-section determines the lowest possible number density of dark  matter today, which in turn determines the highest possible mass, if this single dark matter component saturates the observed abundance. In Fig.~\ref{fig:semi_results2} we impose the unitarity bound on $\langle \sigma v \rangle_S = \langle \sigma v \rangle_{\rm uni}$, where, for s-wave annihilation the unitarity upper bound is given by~\cite{Griest:1989wd, Hui:2001wy}:
\begin{equation}
 \langle \sigma v \rangle_{\rm uni} = (4 \pi/ m_\chi^2)(x_F/ \pi)^{1/2}.
 \label{eq:uni}
 \end{equation}
Here, $x_F=m_\chi/T_F$, with $T_F$ being the freeze-out temperature of the corresponding process. For both the plots in Fig.~\ref{fig:semi_results2}, $x_F=20$ is set as a benchmark value. With the semi-annihilation cross-section set at the s-wave unitarity upper bound, we show the contour in the $m_\chi - \epsilon$ plane (blue solid line) for which in  $\Omega h^2 = 0.12$ is reproduced in the left plot of Fig.~\ref{fig:semi_results2}. This figure shows the maximum possible DM mass allowed for a particular value of $\epsilon$, and as discussed earlier, higher values of $\epsilon$ imply that the upper bound on the DM mass is stronger. In order to understand the approach towards $\epsilon \rightarrow 1$ better, we show in the right panel of Fig.~\ref{fig:semi_results2} a narrower region along the $\epsilon$ axis. From this figure we observe a number of important results:
\begin{enumerate}
\item {\em With the semi-annihilation process alone, one can obtain a scenario giving rise to a nearly complete asymmetry in the DM sector, in which only the (anti-)DM state survives today}. This is obtained for a large value of the CP violation parameter $\epsilon$. Smaller values of $\epsilon$ correspond to scenarios with a mixed present abundance of DM, with both the particle and anti-particle states present.

\item As mentioned above, here we explicitly observe that the approach to $\epsilon \rightarrow 1$ is asymptotic, and correspondingly to $\eta \rightarrow 1$.

\item For $\epsilon \rightarrow 0$, the upper bound on the DM mass is obtained to be $80$ TeV, which is the bound for purely symmetric semi-annihilation scenario, with no CP-violation. 

\item {\em For $\epsilon \rightarrow 1$, the upper bound on the DM mass is obtained to be around $15$ GeV, which is the bound for purely asymmetric semi-annihilation scenario, with maximal CP-violation.} We note that this is much stronger than the unitarity bounds obtained for asymmetric DM scenarios where strong subsequent pair-annihilations are necessarily present, which we consider in the next section~\cite{Baldes:2017gzw}. The above upper bound of $15$ GeV is obtained  for $\eta \simeq 1.0002$, which represents a scenario with a nearly complete present asymmetry in the DM sector (2 particles in 10,000 anti-particles). We have checked that if we reduce $\eta$ further closer to $1$, the consequent change in this upper bound on the DM mass is very small. 

\item We see that being entirely within the limits of maximal possible semi-annihilation rate and the maximal possible value of CP-violation, we can indeed obtain a completely asymmetric DM scenario, with no requirement of subsequent pair-annihilations to remove the symmetric component. This is one of the primary important observations of this paper.
\end{enumerate}

\section{The interplay of semi-annihilation and pair-annihilation}
\label{sec:interplay}
We now consider the second scenario, in which both the semi-annihilation and pair-annihilation processes are active, and their interplay determines the resulting DM properties. In the latter process, a dark matter particle annihilates with an anti-dark matter particle, creating a pair of unstable states $\phi$, $\chi + \chi^\dagger \rightarrow \phi + \phi$, where as earlier $\phi$ can mix with or decay to the SM states. We shall parametrize the leading-order cross-section for this process by $\sigma_A$, which is an additional parameter in this scenario. 

\begin{figure}[htb!]
\label{fig:temp}
\centering
\includegraphics[scale=0.3]{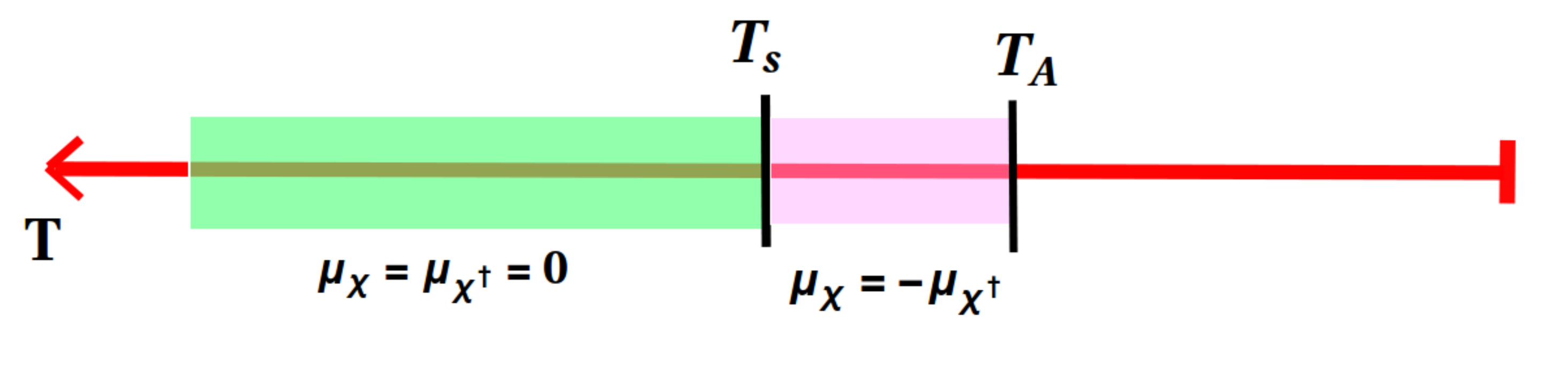}
\caption{\small{\em One of the two possible hierarchies of the freeze-out temperature for the semi-annihilation ($T_S$) and pair-annihilation ($T_A$) processes, and the chemical potentials of the DM ($\mu_\chi$) and anti-DM ($\mu_{\chi^\dagger}$) states at different temperature intervals.}}
\label{fig:temperature}
\end{figure}

We assume that initially at high enough temperatures, both the semi-annihilation and the pair annihilation processes are in chemical equilibrium, with their freeze-out temperatures being $T_S$ and $T_A$ respectively. If the freeze-out temperatures have the hierarchy $T_S > T_A$, the semi-annihilation process freezes out earlier, as schematically shown in Fig.~\ref{fig:temperature}. For temperatures 
$T>T_S > T_A$, using the conditions of detailed balance for the reactions $\chi + \chi \rightarrow \chi^\dagger + \phi$, $\chi + \chi^\dagger \rightarrow \phi+ \phi$ and $\chi^\dagger + \chi^\dagger \rightarrow \chi +\phi$, we obtain the 
relation between the chemical potentials $\mu_\chi = \mu_{\chi^\dagger}=\mu_\phi$. For the cases when $\mu_\phi=0$, this implies that $\mu_\chi = \mu_{\chi^\dagger}=0$.

For $T_A<T < T_S$, the semi-annihilation process freezes out, keeping only the pair annihilation in chemical equilibrium. This would imply that $\mu_\chi + \mu_{\chi^\dagger}=2\mu_\phi$, and if $\mu_\phi=0$ we obtain $\mu_\chi = - \mu_{\chi^\dagger}$. Hence, in this temperature regime, the $\chi$ particle can have a non-zero chemical potential, and therefore, a particle anti-particle asymmetry in the $\chi$ sector is generically possible. Such an asymmetry is generated by the freeze-out of the semi-annihilation process once all the Sakharov conditions are satisfied. In this case, since the pair annihilation process is active for $T<T_S$, the  final particle anti-particle number density ratio ($n^0_\chi/n^0_{\chi^\dagger}$) is determined by both the reaction rates. 

For the opposite hierarchy $T_S < T_A$, there cannot be any chemical potential for the $\chi$ particle for temperatures $T>T_S$, with $\mu_\phi=0$. After the freeze-out of the semi-annihilation, asymmetry may again be generated, as discussed in Sec.~\ref{sec:mi} for the scenario with only semi-annihilation. In particular, since the pair annihilation process is no longer active for $T<T_S$, the ratio ($n^0_\chi/n^0_{\chi^\dagger}$) is entirely determined by the semi-annihilation process. Thus this scenario is identical to the scenario considered in Sec.~\ref{sec:mi} as far as the present DM properties are concerned.

With the pair-annihilation process included in addition to the two CP-conjugate semi-annihilation channels, there are now three relevant processes in the thermal bath that can change the number of DM particles $\chi$. Consequently, the Boltzmann equations~\ref{eq:boltz_semi} are now modified to include an additional collision term as follows:
\begin{align}
\label{eq:boltz_both}
\dfrac{dn_\chi}{dt}+3Hn_{\chi} &= C_{\rm semi}-\int \prod^{4}_{i=1} \frac{d^3 p_i}{(2\pi)^3 2 E_{p_i}}g^2_{\chi}   (2 \pi)^4 \delta^{(4)}(p_1+p_2-p_3-p_4) \nonumber \\ 
& \left[ \overline{|M|^2}_{\chi\chi^{\dagger}\rightarrow \phi\phi}\left[f_{\chi}(p_1)f_{\chi^{\dagger}}(p_2)-f_{\phi}(p_3)f_{\phi}(p_4)\right]
\right],
\end{align}
where $C_{\rm semi}$ is the collision term from the semi-annihilation processes given in the RHS of Eq.~\ref{eq:boltz_semi}, all other notations being the same as in Eq.~\ref{eq:boltz_semi}.

Following the same procedure as in the previous section, the coupled set of Boltzmann equations for the dark matter particle and anti-particle co-moving number densities are now modified to take the following form:
\begin{eqnarray}
\dfrac{d Y_{\chi}}{d x} &=& -\dfrac{s}{H x}\left[A_S\left(Y^2_{\chi}+\dfrac{Y_0 Y_{\chi}}{2}\right)-B_S\left(\dfrac{Y^2_{\chi^{\dagger}}} {2}+Y_0 Y_{\chi^{\dagger}}\right)+\expval{\sigma v}_A \left(Y_{\chi}Y_{\chi^{\dagger}}-Y^2_0\right)\right] \nonumber \\ 
\dfrac{d Y_{\chi^{\dagger}}}{d x} &=& -\dfrac{s}{H x}\left[B_S\left(Y^2_{\chi^{\dagger}}+\dfrac{Y_0 Y_{\chi^{\dagger}}}{2}\right)-A_S\left(\dfrac{Y^2_{\chi}} {2}+Y_0 Y_{\chi}\right)+\expval{\sigma v}_A \left(Y_{\chi}Y_{\chi^{\dagger}}-Y^2_0\right)\right]
\label{boltz1_both}
\end{eqnarray}  
Here, $\expval{\sigma v}_A$ is the thermally averaged pair-annihilation cross-section. As before, we have assumed that throughout the evolution of the $\chi$ and $\chi^\dagger$ particles until the freeze-out of the semi-annihilation and the pair-annihilation processes, the $\phi$ particle is in thermal equilibrium with the SM plasma with a vanishing chemical potential.

\subsection{Estimate of relic abundance}
\label{sec:estimate}
Before proceeding to the discussion of the numerical solutions for the coupled Boltzmann equations, we first provide a rough estimate of the relation between the DM relic density ($\Omega_\chi$), its mass ($m_\chi$), and the CP-violation parameter $\epsilon$.  For this estimate, we shall assume that there is complete asymmetry between the dark matter and anti-matter states in the current Universe, i.e., either only the particle or the anti-particle states survive. It then follows that the present DM relic density $\Omega_{\rm DM} = m_\chi s_0 \left(Y_\chi^\infty +Y_{\chi^\dagger}^\infty \right)/\rho_c =  m_\chi s_0 Y_\chi^\infty /\rho_c$, in the scenario when only the $\chi$ particles survive today, where $s_0$ and $\rho_c$ are the present entropy density and the critical density respectively. 

After the freeze-out of the semi-annihilation process, in the absence of subsequent pair-annihilations, both the $\chi$ and $\chi^\dagger$ co-moving number densities ($Y_\chi $ and $Y_{\chi^\dagger}$) remain constants. However, in the presence of subsequent pair annihilations, namely, the process $\chi \chi^\dagger \rightarrow \phi \phi$, at temperatures below $m_\chi$ (when the backward process is not active), each reaction reduces both $\chi$ and $\chi^\dagger$ numbers by one unit. Therefore, in this latter case, only $Y_\chi - Y_{\chi^\dagger}$ remains a constant, which we can, therefore, equate to $Y_\chi^\infty$, assuming the symmetric part is completely annihilated, and only $\chi$ particles survive today. We now define the net co-moving charge density in the dark matter sector at the temperature $T$ to be $\Delta B (T) = Q \left(n_\chi (T)-n_{\chi^\dagger} (T) \right)$, where $Q$ is  the charge assigned to one DM particle. We can then express the present relic abundance of DM as 
\begin{equation}
\Omega_{\rm DM} = \frac{m_\chi s_0 \Delta B (T_S)}{\rho_c \,s(T_S) Q}, 
\end{equation}
where $T_S$ is the freeze-out temperature for the semi-annihilation process.

In the semi-annihilation reaction, $\chi \chi \rightarrow \chi^{\dagger} \phi$, the net change in $\chi$ charge per reaction is negative ($\Delta Q=-3Q$), while in the CP-conjugate process $\chi^{\dagger} \chi^{\dagger}\rightarrow \chi \phi$, the net change in $\chi$ charge per reaction is positive ($\Delta Q=3Q$). Hence, the probability of having a positive change is $P_{+} = \sigma_{\bar{S}} / \left(\sigma_S + \sigma_{\bar{S}}\right)$, while the probability for a negative change is $P_{-} =\sigma_S / \left(\sigma_S + \sigma_{\bar{S}}\right)$. Therefore, $\Delta Q$ produced per semi-annihilation and its CP-conjugate reaction is $(3QP_{+} -3QP_{-})=-3Q\epsilon$, where $\epsilon$ is defined as in Eq.~\ref{eq:epsilon}. Here, we have used the fact that the final state phase space elements are the same for the two CP-conjugate processes. Finally, the net DM charge density produced is $\Delta B = - 3 \epsilon Q n_{\chi}^{\rm eq} (T_S)$, 
assuming that the near-equilibrium distribution with zero chemical potential, $n_{\chi}^{\rm eq} (T_S)$, is being maintained by fast pair-annihilation reactions, and therefore, $n_{\chi}^{\rm eq} (T_S) \simeq n_{\chi^\dagger}^{\rm eq} (T_S)$.
Since this assumption is invalid for the scenario discussed in Sec.~\ref{sec:mi}, our estimate of the relic abundance in this section does not apply for that scenario. In particular, with only the CP-violating semi-annihilation reaction active in the thermal bath, for large CP-violation (which is necessary to get a complete asymmetry with only semi-annihilation) the DM and the anti-DM particles have large and different pseudo-chemical potentials, and therefore do not follow the equilibrium distribution.

Plugging in the expression for $\Delta B$ as obtained above, we can now write the relic abundance of DM particles today as follows
\begin{align}
\Omega_{\rm DM} &= \frac{3~m_\chi s_0 |\epsilon| ~n_\chi^{\rm eq} (T_S)}{\rho_c ~ s(T_S) } ~~~{\rm (completely ~asymmetric ~scenario)} .
\label{eq:estimate}
\end{align}
Since we assumed the DM state $\chi$ to survive in the present Universe, $\epsilon<0$ in this case, while if the anti-DM state $\chi^\dagger$ survives, $\epsilon>0$, as can be seen from Eq.~\ref{eq:epsilon}. We can re-write the above expression in terms of a set of particular choices of the parameters as
\begin{equation}
\Omega_{\rm DM}h^2 = 0.12 \times ~ \bigg(\frac{m_\chi}{100 ~\rm GeV}\bigg) \bigg(\frac{100}{g_{*S}(x_F)}\bigg) ~ \bigg( \frac{|\epsilon| ~x_S^{3/2}~ e^{-x_S}}{10^{-9}} \bigg).
\label{eq:estimate2}
\end{equation}
This shows that apart from the implicit dependence of $\Omega_{\rm DM}$ on $m_\chi$ and $\epsilon$ through the value of $x_S(=m_\chi /T_S)$, there is an explicit linear proportionality with both these parameters expected. This is to be contrasted with the simple scenarios of asymmetry generation through the out-of-equilibrium decay of a heavy particle, where the resulting particle density today is proportional to $\epsilon$ only, and not to the mass of the decaying heavy particle~\cite{Kolb:1990vq}. Furthermore, in the decay scenario, the asymmetry parameter $\epsilon$ is independent of the particle momenta, unlike in the annihilation scenario~\cite{Kolb:1990vq}. For a typical value of $x_S = 20$, we see that $|\epsilon| \simeq 5.4 \times 10^{-3}$ can reproduce the present DM abundance, for $m_\chi = 100$ GeV.  {\em In contrast to the scenario with only semi-annihilation discussed in Sec.~\ref{sec:mi}, we see that the CP-violation required to generate complete asymmetry here is very small.}

Unlike in the previous case, for pair-annihilation cross-sections that are not sufficient to completely remove the symmetric component, there is an explicit dependence of the DM relic density on the pair-annihilation rate $\langle \sigma v \rangle_A$. In this case, the coupled Boltzmann equations can be integrated piecewise in different temperature regimes, firstly near the freeze-out of the semi-annihilation process, in which the pair-annihilation rate is not relevant, and then near the freeze-out of the pair-annihilation process, but now with an initial asymmetry in the DM sector generated by the earlier freeze-out of the semi-annihilation. The resulting relic abundance can then be expressed as~\cite{Graesser:2011wi, Iminniyaz:2011yp}:
\begin{equation}
\Omega_{\rm DM} = \frac{s_0}{\rho_c}m_{\chi} C \coth \left(\frac{C\lambda \langle \sigma v \rangle_A} {2 x_A} \right) ~~~~{\rm (partially ~asymmetric ~scenario),}
\end{equation}
where, $x_A=m_{\chi}/T_A$, with $T_A$ being the freeze-out temperature of the pair-annihilation process,  $C=Y_\chi (T)- Y_{\chi^\dagger}(T)$, for all $T<T_S$, and $\lambda=1.32 M_{\rm Pl}m_{\chi} g^{1/2}_*$. In the limit $C \rightarrow 0$, the above expression reduces to the well-known result for symmetric WIMP scenario
\begin{equation}
\Omega_{\rm DM}= \frac{2 s_0 m_{\chi} x_A}{\rho_c \lambda \langle \sigma v \rangle_A} ~~~~~~{\rm (symmetric ~WIMP ~scenario)}.
\end{equation}

\subsection{Numerical results}
We shall now solve the coupled Boltzmann equations~\ref{boltz1_both} numerically, with four free parameters. The three parameters $m_\chi$, $\langle \sigma v \rangle_S$ and $\epsilon_{\rm eff}$ are the same as in Sec.~\ref{sec:mi}, with the additional parameter being the pair-annihilation rate $\langle \sigma v \rangle_A$. Since we have already discussed the role of the first three parameters in determining the DM properties in the previous section, the primary aim of this section is to understand the impact of pair-annihilation, in particular its interplay with the semi-annihilation process. Following our general discussion above, therefore, the relevant temperature hierarchy is $T_S > T_A$, in which the semi-annihilation freezes out earlier. The opposite hierarchy, $T_S < T_A$, is exactly equivalent to the scenario in Sec.~\ref{sec:mi}, as far as the DM asymmetry and relic density today are concerned.

\begin{figure}[htb!]
\begin{center}
\includegraphics[scale=0.6]{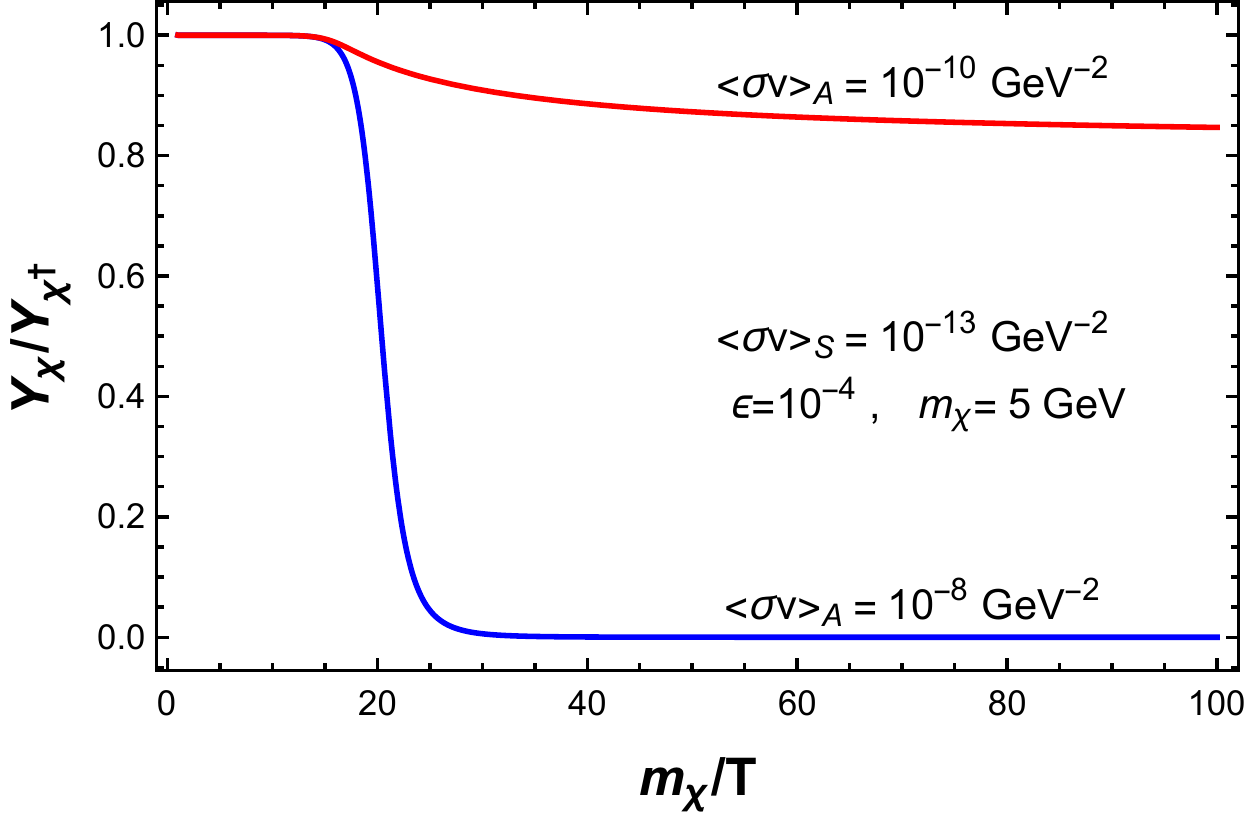} 
\caption{\small{\em Variation of the ratio of dark matter and anti-dark matter yields, $Y_\chi / Y_{\chi^\dagger}$, as a function of $m_\chi/T$, for different values of the pair-annihilation rate, $\langle \sigma v \rangle_A=10^{-10} {~\rm GeV}^{-2}$ (red solid line) and $\langle \sigma v \rangle_A=10^{-8} {~\rm GeV}^{-2}$ (blue solid line), with fixed values of $m_\chi$, $\langle \sigma v \rangle_S$ and $\epsilon$.}}
\label{fig:Yield_pair}
\end{center}
\end{figure}
For  $T_S > T_A$, the essential role of the pair-annihilation process is to remove the symmetric component of dark matter, as illustrated in Fig.~\ref{fig:Yield_pair}. As we can see from this figure, with fixed values of $m_\chi$, $\langle \sigma v \rangle_S$ and $\epsilon$, if we increase the value of the pair-annihilation rate from $\langle \sigma v \rangle_A=10^{-10} {~\rm GeV}^{-2}$ (red solid line) to $\langle \sigma v \rangle_A=10^{-8} {~\rm GeV}^{-2}$ (blue solid line), the ratio of dark matter and anti-dark matter yields $Y_\chi / Y_{\chi^\dagger}$ decreases rapidly.

\begin{figure}
\includegraphics[scale=0.52]{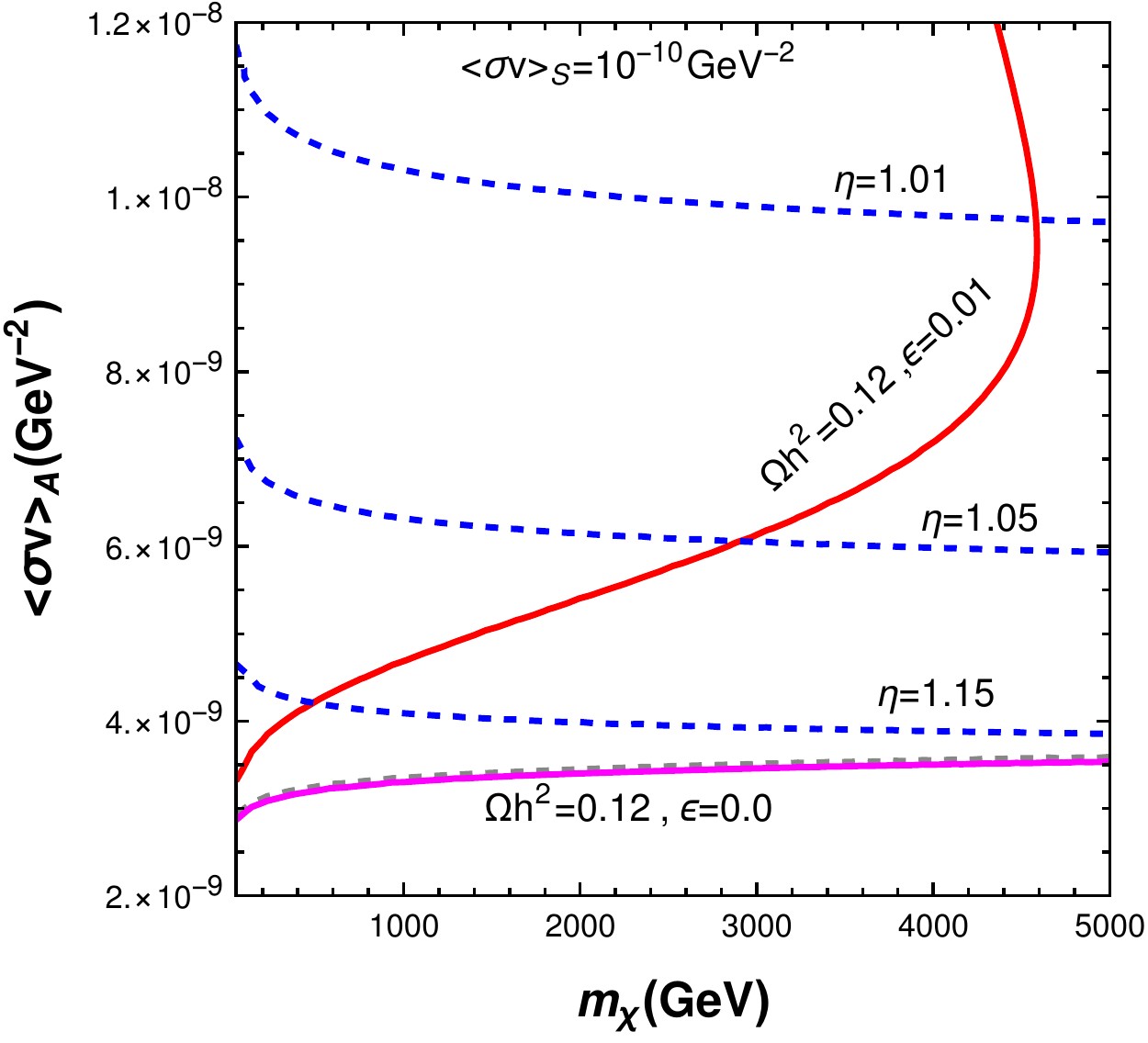} \hspace*{1.5cm}
\includegraphics[scale=0.52]{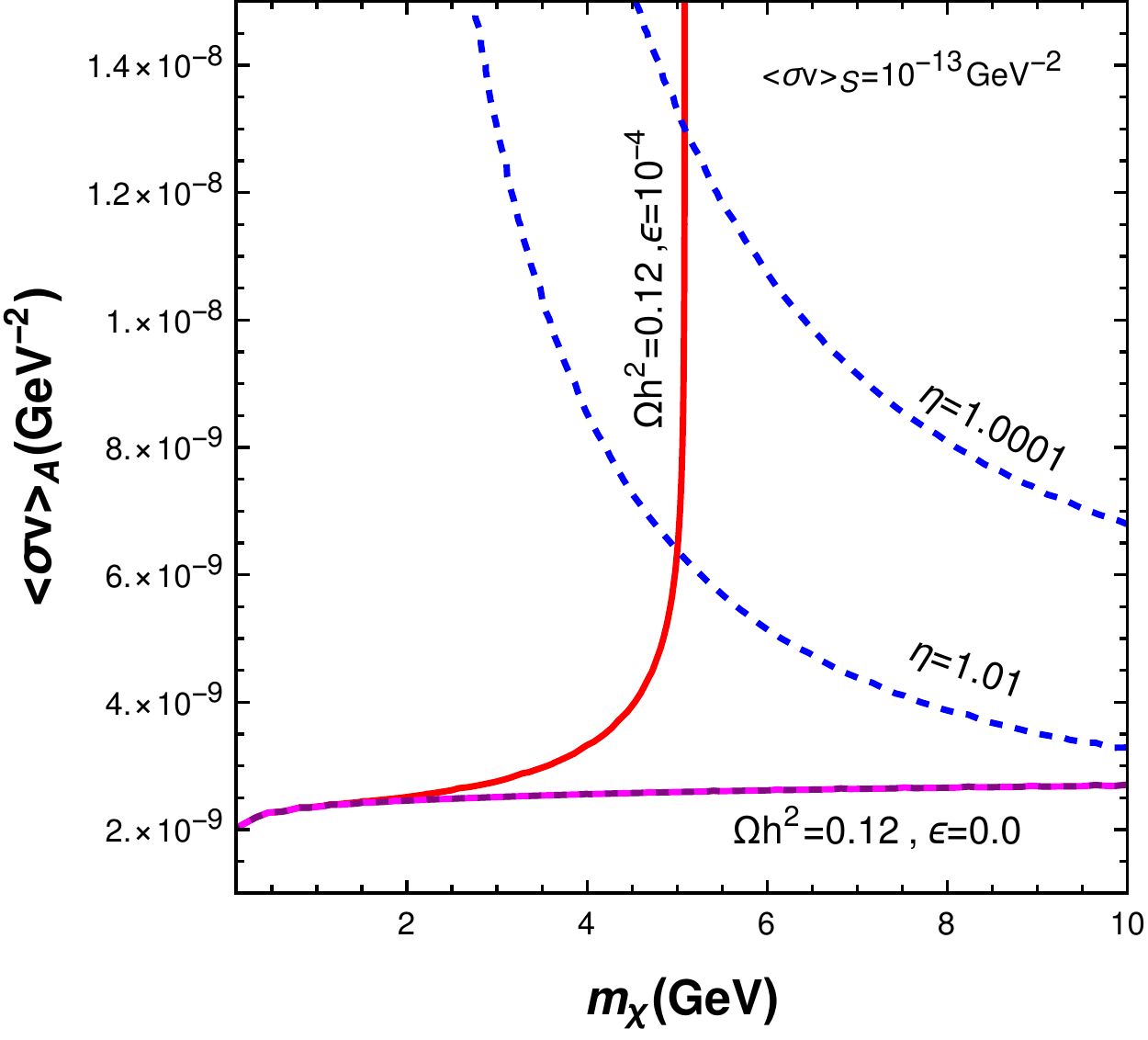}
\caption{\small{\em Required values of the pair-annihilation cross-section $\langle \sigma v \rangle_A$ as a function of the DM mass $m_\chi$ that can reproduce the observed DM relic abundance (red solid line). The results are shown for different values of the CP-violation parameter $\epsilon$ and the semi-annihilation rate $\langle \sigma v \rangle_S$ (left and right plots). Corresponding results for the symmetric case ($\epsilon=0$) are also shown for comparison, with semi-annihilation (black dashed line) and without it (pink solid line), the latter case being the pure WIMP scenario. Also shown are the contours for the present DM relative abundance parameter $\eta$. See text for details.}}
\label{Fig:param1}
\end{figure}

In order to understand the typical values of the cross-sections required to reproduce the observed relic abundance today, we show in Fig.~\ref{Fig:param1} the regions in the $\langle \sigma v \rangle_A$ and $m_\chi$ parameter space in which the central value of the DM relic density  $\Omega h^2 = 0.12$ is reproduced. For both the plots in this figure (left and right), the values of $\epsilon$ and $\langle \sigma v \rangle_S$ have been kept fixed. We show the results for $\epsilon=0.01$ and $\langle \sigma v \rangle_S = 10^{-10} {~\rm GeV}^{-2}$ in the left figure, and for $\epsilon=10^{-4}$ and $\langle \sigma v \rangle_S = 10^{-13} {~\rm GeV}^{-2}$ in the right figure. We also show contours in the $m_\chi - \langle \sigma v \rangle_A$ parameter space for constant values of the present relative abundance of DM and anti-DM, parametrized by $\eta$, as defined in Eq.~\ref{eq:eta}. 

As expected from our discussion in Sec.~\ref{sec:estimate}, in particular Eq.~\ref{eq:estimate2}, as we increase $\epsilon$, the value of $\langle \sigma v \rangle_S$ required to reproduce the relic density is also correspondingly increased. This is primarily due to the exponential suppression of $\Omega h^2$ from $x_S$, which is increased for larger $\langle \sigma v \rangle_S$, thereby requiring larger $\epsilon$. This estimate is applicable only in the case when the symmetric component today is negligible, i.e., $\eta \simeq 1$. As we see in Fig.~\ref{Fig:param1}, in the entire parameter space under consideration, $\eta$ is close to $1$. We also show the contour for $\Omega h^2 = 0.12$, in the $\epsilon=0$ limit (black dashed line), which is found to overlap with the corresponding contour (pink solid line) in the case in which only the pair-annihilation is active (i.e., $\langle \sigma v \rangle_S=0$ as well). This is not surprising, since for such small values of $\langle \sigma v \rangle_S$, which is at least an order of magnitude below the values of $\langle \sigma v \rangle_A$, semi-annihilation is essentially not relevant in determining the present DM abundance as long as $\epsilon=0$. 

The scenario, however, changes dramatically with the introduction of a small CP-violation with a non-zero $\epsilon$, when semi-annihilation becomes the key process in determining the present density. The role of $\langle \sigma v \rangle_A$ for non-zero $\epsilon$ is then to eliminate the symmetric component of DM that is left over at the freeze-out of the semi-annihilation process. As we have already seen in Sec.~\ref{sec:mi}, for large $\mathcal{O}(1)$ values of $\epsilon$, no other number changing process plays any role in determining the relic abundance. This is because such a scenario leads to a large violation of CP in the DM sector, thereby producing an almost completely asymmetric DM already at the freeze-out of the semi-annihilation process at temperature $T_S$. Since almost no symmetric component is left out in this case at $T=T_S$, the pair-annihilation process is not relevant. 

In the limit $\eta \rightarrow 1$, we see from Eq.~\ref{eq:estimate2} that for a fixed value of $x_S$ (which in turn is obtained for a fixed value of $\langle \sigma v \rangle_S$ in this limit) and $\epsilon$, the dark matter mass is also fixed. In particular, as we see from Fig.~\ref{Fig:param1}, with $\epsilon=0.01$ and $\langle \sigma v \rangle_S = 10^{-10} {~\rm GeV}^{-2}$, we obtain $m_\chi \sim 4600$ GeV, while for $\epsilon=10^{-4}$ and $\langle \sigma v \rangle_S = 10^{-13} {~\rm GeV}^{-2}$ , $m_\chi \sim 5$ GeV. Away from the region in the parameter space for which $\eta \rightarrow 1$, we find it non-trivial to obtain a semi-analytic solution to the Boltzmann equations. However, it is clear from Fig.~\ref{Fig:param1} that the DM mass is no longer uniquely fixed for such a case, but varies with $\langle \sigma v \rangle_A$. This is essentially because the symmetric component is not completely removed in such scenarios.

We note in passing that the parameter values $\epsilon=10^{-4}$ and $\langle \sigma v \rangle_S = 10^{-13} {~\rm GeV}^{-2}$ predict a DM mass of around $5$ GeV in the completely asymmetric DM limit. Since this value of the DM mass is around five times the proton mass, we expect the current number densities of the surviving DM particle and protons to be similar in this scenario. As is well-known, such a DM mass is also expected in scenarios which dynamically relate the DM and baryon number densities in the current Universe~\cite{Petraki:2013wwa, Zurek:2013wia}. Such a mechanism to relate the two asymmetries might be possible through semi-annihilation.

In the pure WIMP scenario, with $\epsilon=0$ and $\langle \sigma v \rangle_S=0$, in the freeze-out approximation, the dependence of $\Omega h^2$ on the DM mass is logarithmic, while it is inversely proportional to $\langle \sigma v \rangle_A$. Therefore, we see in Fig.~\ref{Fig:param1} that the value of $\langle \sigma v \rangle_A$ required (around $3.5 \times 10^{-9} {~\rm GeV}^{-2}$) to reproduce $\Omega h^2=0.12$ is largely independent of $m_\chi$ (pink solid line in both figures). As discussed above, this value remains unchanged with the introduction of a small $\langle \sigma v \rangle_S$, when the CP-violation is zero ($\epsilon=0$). In the $\eta \rightarrow 1$ limit, for non-zero $\epsilon$, the requirement of $\langle \sigma v \rangle_A$ is larger, and it increases with increasing $\epsilon$. 

\begin{figure}[htb!]
\centering
\includegraphics[scale=0.52]{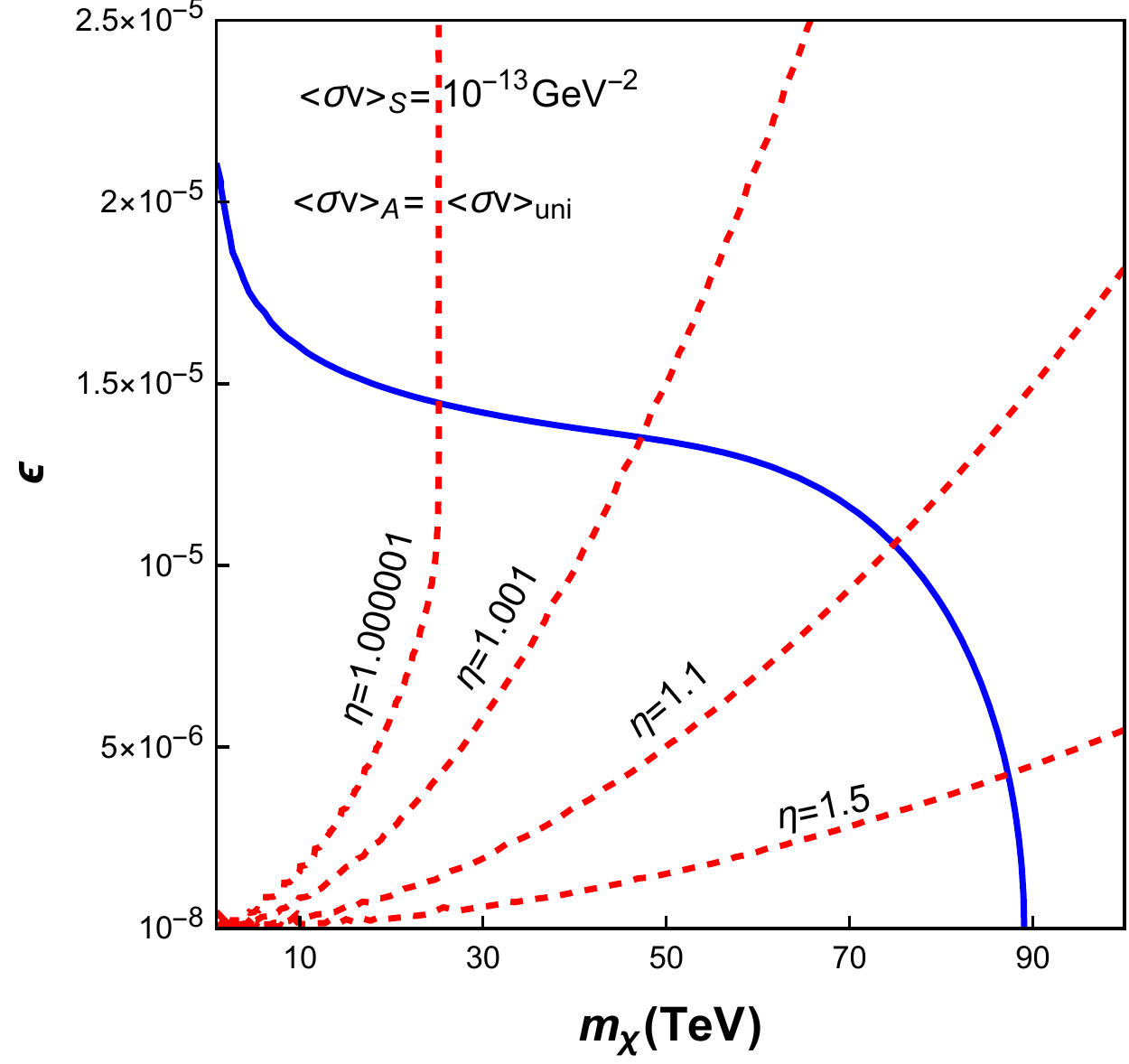}
\caption{\small{\em Contours in the $m_\chi - \epsilon$ plane in which the central value of the DM relic density  $\Omega h^2 = 0.12$ is reproduced, with the pair-annihilation rate fixed at its s-wave upper bound implied by S-matrix unitarity, $\langle \sigma v \rangle_A = \langle \sigma v \rangle_{\rm uni} = (4 \pi/ m_\chi^2)(x_F/ \pi)^{1/2}$. The semi-annihilation cross-section is fixed to ensure $ \langle \sigma v \rangle_{S} <  \langle \sigma v \rangle_A$, for all values of $m_\chi$ considered in this figure, such that the freeze-out temperature hierarchy $T_S > T_A$ is satisfied.}}
\label{fig:uni}
\end{figure}

As in Sec.~\ref{sec:mi}, we can obtain an upper bound for possible values of the dark matter mass by using partial-wave unitarity to bound the annihilation (or semi-annihilation) cross-sections from above. For the scenario in which $T_S>T_A$, the annihilation cross-section must be larger than the semi-annihilation cross-section, and therefore, we impose the unitarity bound on $\langle \sigma v \rangle_A = \langle \sigma v \rangle_{\rm uni}$, where, $\langle \sigma v \rangle_{\rm uni}$ is as given in Eq.~\ref{eq:uni}. In this case, we show the resulting upper bound on the dark matter mass as a function of the CP-violation parameter $\epsilon$ in Fig.~\ref{fig:uni}. We have fixed the value of the semi-annihilation cross-section to be $ \langle \sigma v \rangle_S = 10^{-13} {~\rm GeV}^{-2}$, which is chosen to ensure that  $ \langle \sigma v \rangle_{S} <  \langle \sigma v \rangle_A =  \langle \sigma v \rangle_{\rm uni}$, for all values of $m_\chi$ considered in this figure.

In Fig.~\ref{fig:uni}, the observed relic abundance $\Omega h^2 = 0.12$ is satisfied along the solid blue line. As in Fig.~\ref{fig:semi_results2}, we see that as the CP-violation parameter $\epsilon$ decreases, the resulting mass bound becomes stronger. Furthermore, higher values of $\epsilon$ lead to larger present asymmetry in the dark matter sector, and therefore a value of $\eta$ closer to $1$. The general result obtained in Sec.~\ref{sec:mi}  that the bound on $m_\chi$ for asymmetric DM is stronger compared to the symmetric DM scenario, continues to hold in this scenario as well. In the complete asymmetric limit, i.e., $\eta \rightarrow 1$, the upper bound on the DM mass is found to be around $25$ TeV (numerically for $\eta=1.000001$), while for $\eta \rightarrow 2$ it's around $90$ TeV, assuming s-wave annihilation. 
For the opposite hierarchy of the freeze-out temperatures, i.e., $T_S<T_A$, the semi-annihilation cross-section must be larger than the pair-annihilation cross-section, and therefore, the unitarity bound must be imposed on $\langle \sigma v \rangle_S$, which has already been discussed in Sec.~\ref{sec:mi}, in particular in Fig.~\ref{fig:semi_results2}.

\begin{figure}[htb!]
\begin{center}
\includegraphics[scale=0.52]{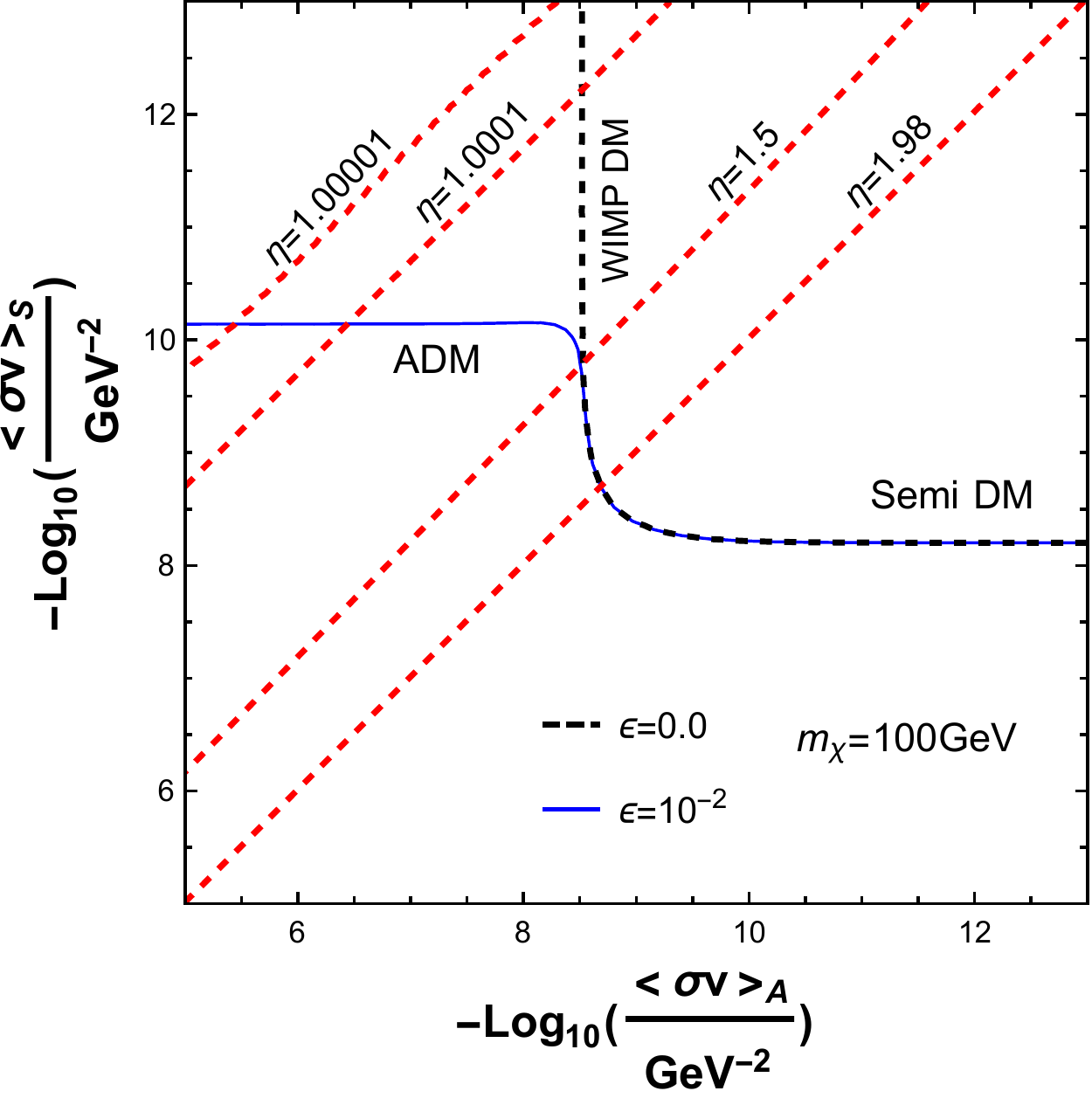} 
\caption{\small{\em Phase diagram showing the interplay between the semi-annihilation and pair-annihilation rates in determining the asymmetric DM properties. The observed relic abundance $\Omega h^2 = 0.12$ is satisfied, with and without CP-violation, along the blue solid and black dashed lines, respectively. The relative abundance parameter $\eta \rightarrow 2$ represents the symmetric phase, while $\eta \rightarrow 1$ represents the asymmetric phase. Here, ADM, Semi DM and WIMP DM denote the dominantly asymmetric dark matter, symmetric semi-annihilating DM and symmetric pair-annihilating DM phases, respectively. See text for details.}}
\label{Fig:param2}
\end{center}
\end{figure}
We can summarize our discussion of the interplay between the semi-annihilation and pair-annihilation rates in determining the asymmetric DM properties using an instructive phase diagram, as shown in Fig.~\ref{Fig:param2}. In this figure, we study the values of $\langle \sigma v \rangle_S$ and $\langle \sigma v \rangle_A$ for which the observed relic abundance $\Omega h^2 = 0.12$ is satisfied, with and without CP-violation. When the CP-violation vanishes, i.e., with $\epsilon=0$, the relic abundance is satisfied along the black dashed contour~\cite{DEramo:2010keq}. Since for $\epsilon=0$, both semi-annihilation and pair-annihilation can reproduce the observed relic abundance, with either or both of them contributing, we obtain an approximate upper bound of $10^{-8} {~\rm GeV}^{-2}$ for both the rates, for a fixed DM mass of $m_\chi=100$ GeV. In contrast, when CP-violation is turned on, i.e., for $\epsilon=10^{-2}$ in Fig.~\ref{Fig:param2}, a symmetric phase and  an asymmetric phase appear in which the relic density is satisfied, as seen in the solid blue line. The two phases can be distinguished by constant values of the DM relative abundance parameter $\eta$. The symmetric phase, with $\eta \rightarrow 2$ is identical to the $\epsilon=0$ scenario, and hence the blue solid line and the black dashed lines overlap. In this phase, the pair-annihilation rate is not large enough to remove the symmetric component efficiently. On the other hand, the asymmetric phase appears when $\langle \sigma v \rangle_A$ is larger than the previously obtained upper bound of around $10^{-8} {~\rm GeV}^{-2}$, for $m_\chi=100$ GeV. In contrast, $\langle \sigma v \rangle_S$ is much smaller in this phase. Thus, to summarize, there are two ways to produce asymmetric DM in the absence of any wash-out processes, namely, 
\begin{enumerate}
\item have a large CP-violation $\epsilon$ as in Sec.~\ref{sec:mi}, in which case semi-annihilation is sufficient to create a complete DM asymmetry, and no subsequent number changing process is necessary, or, 

\item produce a small asymmetry through a small CP-violation $\epsilon$, and then have a sufficiently large pair-annihilation rate to remove the symmetric component, as shown in this section, and as is clear from Fig.~\ref{Fig:param2}.
\end{enumerate}

\section{Complex scalar DM with cubic self-interaction}
\label{model}
We now discuss a simple toy model in which the generic scenario described in Sec.~\ref{sec:interplay} with both the semi- and pair-annihilation processes can be realized. The minimal new field content that can lead to a particle-antiparticle asymmetry through the semi-annihilation process include a complex scalar $\chi$ which is charged under a $Z_3$ symmetry (we assign the charge $\omega$ to $\chi$, where $\omega^3=1$) and a real scalar $\phi$, which is a singlet under this symmetry, as well as the SM gauge interactions. The SM fields are also singlets under the discrete $Z_3$ symmetry. The $Z_3$ symmetry ensures the stability of $\chi$, making it the DM candidate. For earlier studies involving different aspects of $Z_3$ symmetric DM, see, for example, Refs.~\cite{Belanger:2012vp, Belanger:2012zr, Hochberg:2014dra, Bernal:2015bla, Hektor:2019ote, Choi:2015bya, Choi:2016tkj}. The effective low-energy interaction Lagrangian involving the $\chi$ and the $\phi$ particles is given by
\begin{equation}
\mathcal{L} \supset \frac{1}{3!} \left(\mu \chi^3 + {\rm h.c.} \right) + \frac{1}{3!} \left(\lambda \chi^3 \phi + {\rm h.c.} \right) + \frac{\lambda_1}{4} \left(\chi^\dagger \chi \right)^2 +   \frac{\lambda_2}{2} \phi^2 \chi^\dagger \chi + \mu_1 \phi  \chi^\dagger \chi +  \frac{\mu_2}{3!} \phi^3 +  \frac{\lambda_3}{4!} \phi^4.
\label{eq:lag}
\end{equation}
Here, the couplings $\mu$ and $\lambda$ can be complex in general. However, one of the phases can be rotated away by an appropriate re-definition of the field $\chi$. Therefore, in this general effective low-energy theory, there is one residual complex phase, which is necessary to generate a CP-asymmetry in the DM sector. We take $\mu$ to be real, and $\lambda$ to have a non-zero imaginary part, with a phase $\theta$. The parameters in the scalar potential in Eq.~\ref{eq:lag} can be suitably chosen to ensure that the $\chi$ field does not develop a VEV, thereby ensuring that the $Z_3$ symmetry is not spontaneously broken.

In addition to the interaction terms involving the $\chi$ and the $\phi$ fields in Eq.~\ref{eq:lag}, there can be two dimension-four and one dimension-three couplings to the SM Higgs doublet $H$ as well, namely, $\lambda_{H\chi} ~(\chi^\dagger \chi  |H|^2) + \lambda_{H \phi} ~(\phi^2 |H|^2) + \mu_{H \phi} ~(\phi |H|^2)$. For $m_{\chi}> m_H$, the $\lambda_{H\chi}$ term contributes in exactly the same way as the $\lambda_2$ term in Eq.~\ref{eq:lag}, and therefore we do not consider it separately here. Furthermore, the $\lambda_{H \phi}$ and $\mu_{H \phi}$ terms lead to interactions of the $\phi$ field with the $H$ field, which will thermalize the $\phi$ field with the SM plasma. Since we assume the $\phi$ particles to be in equilibrium with the SM bath with zero chemical potential, the effect of these terms are also included.

The above interaction Lagrangian in Eq.~\ref{eq:lag} leads to several class of $2 \rightarrow 2$, $2 \rightarrow 3$ and $3 \rightarrow 2$ processes. We find that in different regions of the multi-dimensional parameter space, different class of diagrams (or combinations thereof) may dominate. Since in this section we are presenting a toy model that realises the general features of the model-independent setup discussed in the previous section, we shall focus on a restricted region of the parameter space in which a subset of the $2 \rightarrow 2$ diagrams dominate. In particular, we shall consider the values of the dimensionful parameters to be small compared to the DM mass scale, i.e., $\mu / m_\chi << 1$ and $\mu_1 / m_\chi << 1$. We shall also take the cubic and quartic self-couplings of the $\phi$ field to be small, which does not alter the qualitative features of the scenario. A comprehensive study of the above toy model will be presented elsewhere~\cite{GGM2}. 

\begin{figure}[htb!]
\centering
\includegraphics[scale=0.55]{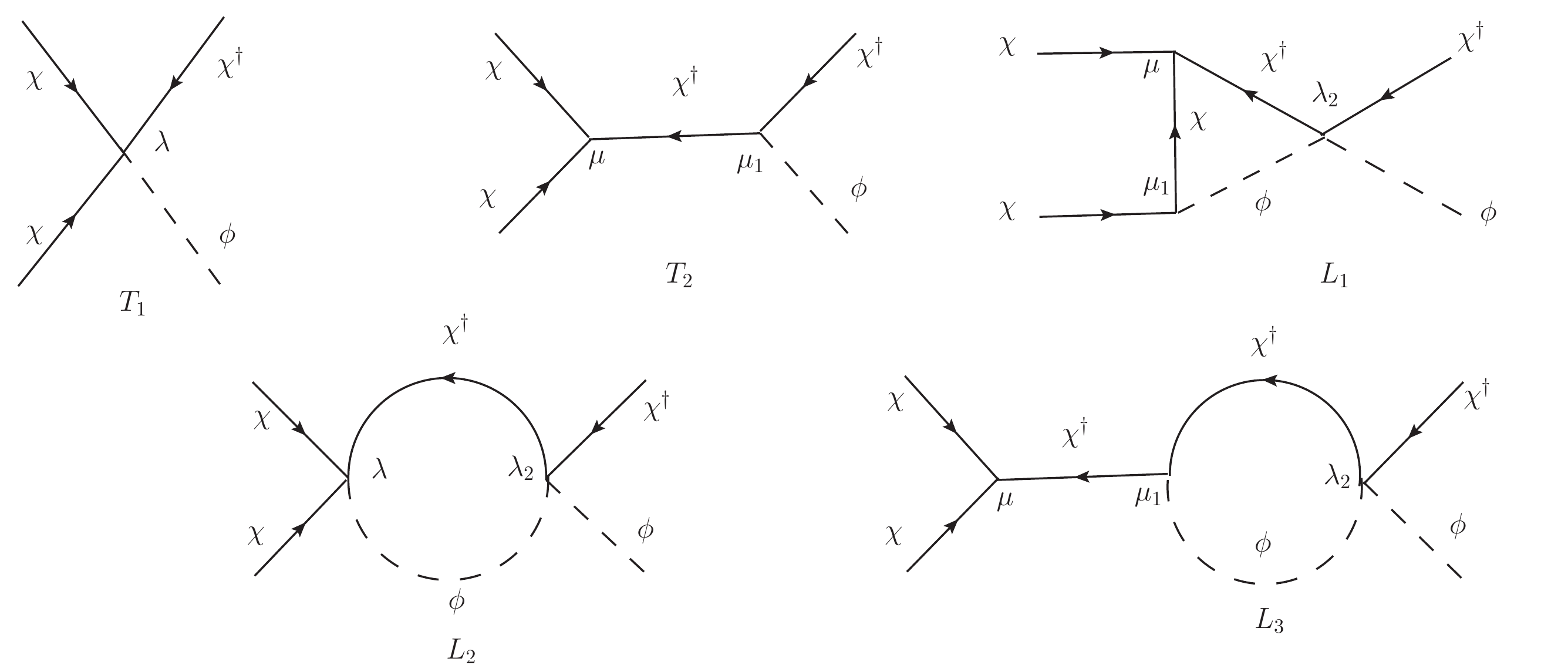}
\caption{\small{\em  Relevant $2 \rightarrow 2$ tree-level and one-loop Feynman diagrams for the semi-annihilation process $\chi \chi \rightarrow \chi^\dagger \phi$.}}
\label{Fig:diagrams}
\end{figure}
The relevant tree-level and one-loop Feynman diagrams for the semi-annihilation process $\chi \chi \rightarrow \chi^\dagger \phi$ are shown in Fig.~\ref{Fig:diagrams}. At tree-level there are two Feynman diagrams contributing to this process: one involving a contact interaction (diagram $T_1$), and the other with an intermediate $\chi$ propagator (diagram $T_2$). The second diagram gives a contribution to the matrix element proportional to $\left (\mu \mu_1/ m_\chi^2 \right)$, in the non-relativistic limit for the $\chi \chi$ initial state, with the centre of mass energy squared $s \simeq 4 m_\chi^2$. Therefore, for  $\mu / m_\chi << 1$ and $\mu_1 / m_\chi << 1$, the contact interaction dominates.

In order to determine the CP-asymmetry generated by the semi-annihilation process, we compute the interference between the tree-level and loop-level diagrams shown in Fig.~\ref{Fig:diagrams}. In general the CP-asymmetry is proportional to  $\Im \left({M}_{\rm tree} (g_i)^* {M}_{\rm loop} (g_j)\right)$, which in turn is proportional to $\Im \left( \prod_{i,j} g_i^* g_j \right) \times \Im ( I)$, where $I$ is the loop factor which acquires an imaginary part when the particles in the loop go on-shell. The latter requirement is ensured by the condition $m_{\phi} < m_{\chi}$. We find that diagram $T_2$ gives a non-zero contribution to the CP-asymmetry, resulting from its interference with the loop diagram $L_2$, while diagram $T_1$ leads to a non-zero contribution from its interference with $L_1$ and $L_3$. Furthermore, the contributions from the interference of $T_2$ and $L_2$, and that from $T_1$ and $L_3$ cancel identically. Therefore, the only relevant contribution is from the interference of $T_1$ and $L_1$. The resulting difference in matrix elements squared that contribute to $\epsilon$ as defined in Eq.~\ref{eq:epsilon}, is given as:
\begin{equation}
|{M}|^2_{\chi\chi\rightarrow \chi^{\dagger} \phi}-|{M}|^2_{\chi^{\dagger}\chi^{\dagger}\rightarrow \chi \phi} = \frac{4 |\lambda| \mu \mu_1 \lambda_2 \sin{\theta}}{16 \pi \sqrt{s(s-4m^2_{\chi})}} \log \left[ \frac{\mdm^2+m^2_{\phi}-s+\beta_1}{\mdm^2+m^2_{\phi}-s-\beta_1}\right] 
\end{equation}
where, $s$ is the centre of mass energy squared, and 
\begin{equation}
\beta_1 =\sqrt{\frac{(s-4\mdm^2)(\mdm^4+(m^2_{\phi}-s)^2-2\mdm^2(s+m^2_{\phi}))}{s}}.  
\end{equation}

\begin{figure}
\includegraphics[scale=0.55]{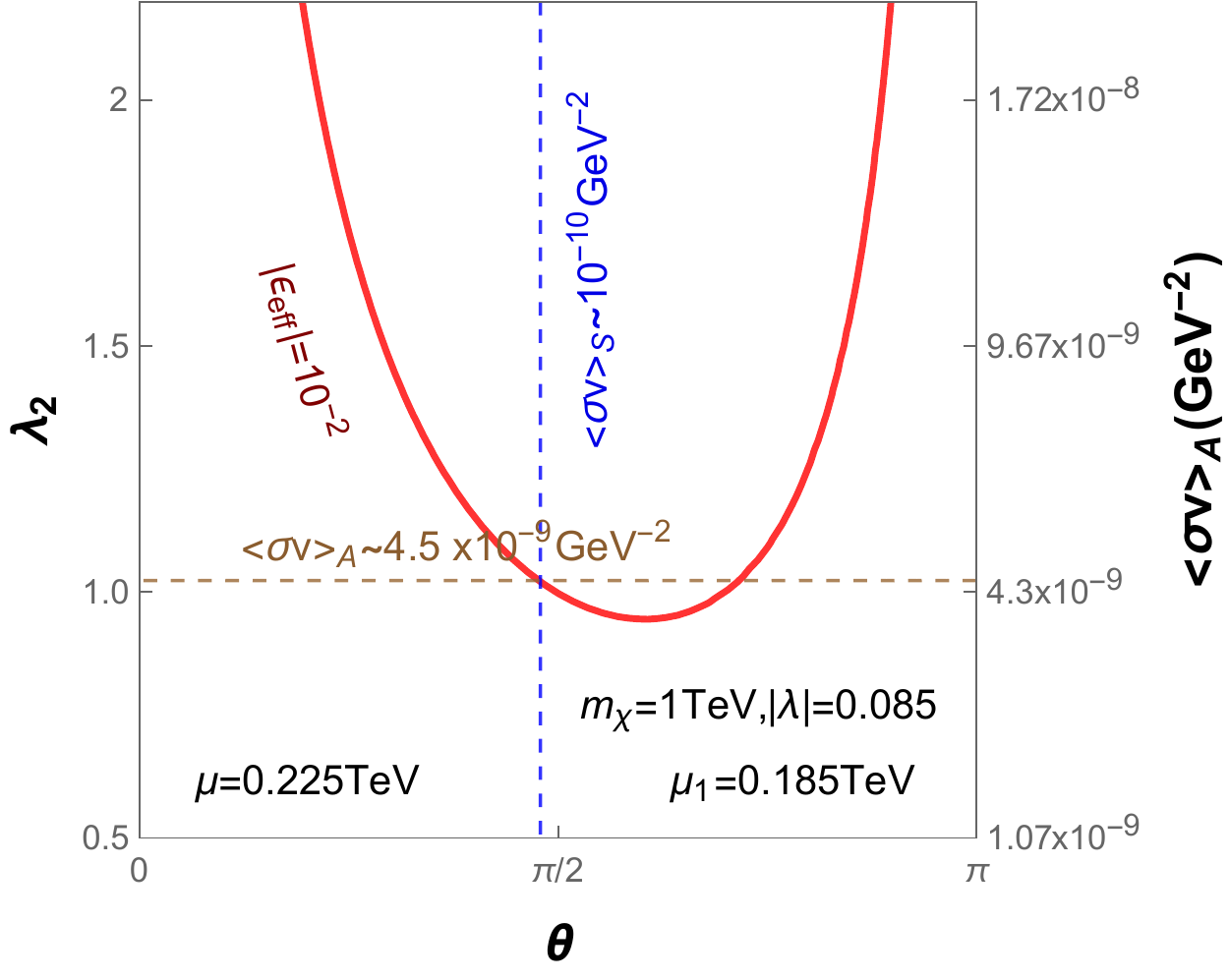} \hspace*{1.0cm}
\includegraphics[scale=0.55]{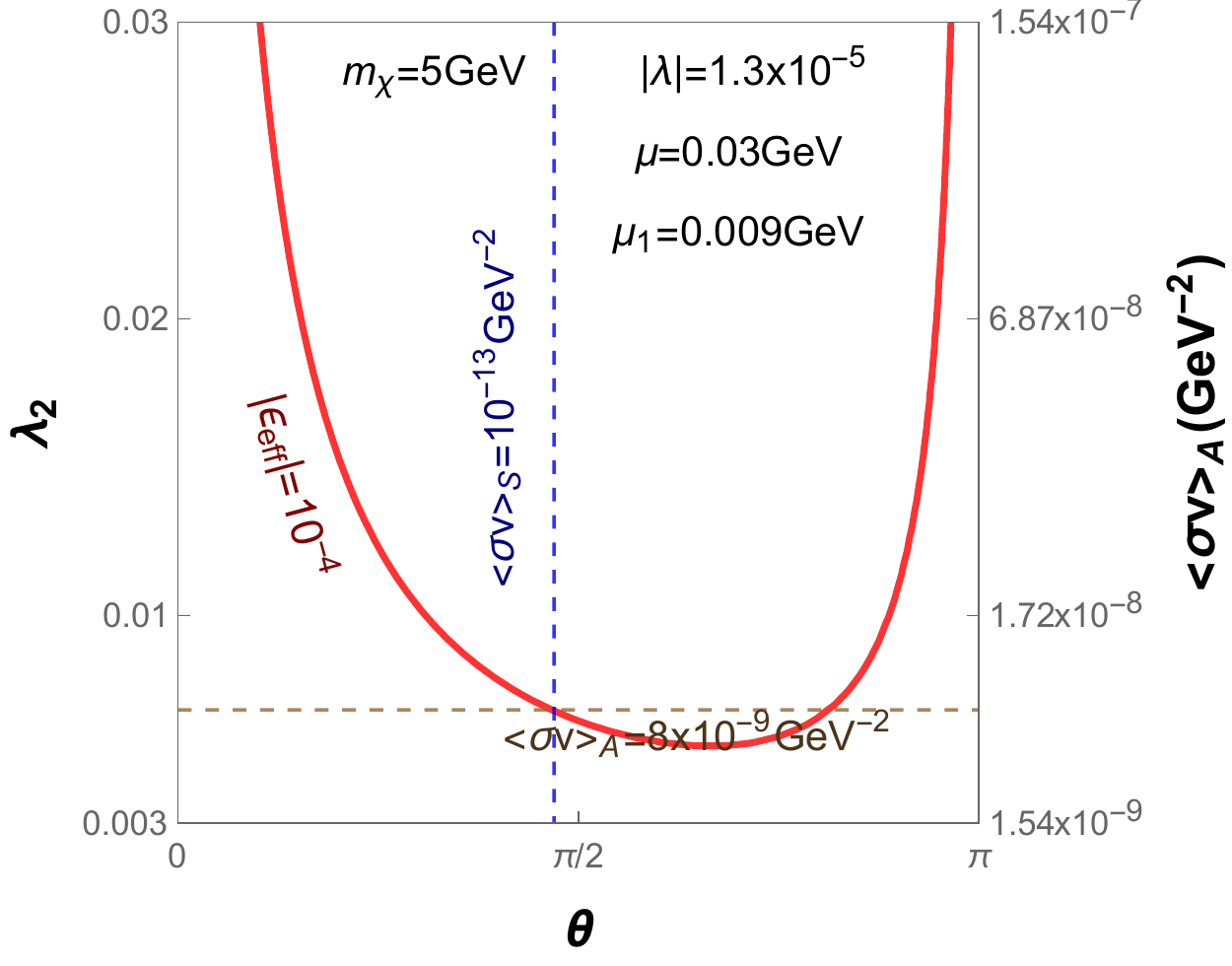}
\caption{\small{\em Contours of fixed effective CP-violation parameter $|\epsilon_{\rm eff}|$ (red solid lines), as a function of the complex phase $\theta=\arg(\lambda)$ and the effective pair-annihilation coupling $\lambda_2$. The results are shown for two different DM mass values $m_\chi=1$ TeV (left panel) and $5$ GeV (right panel). The required values of the annihilation rates and $\epsilon_{\rm eff}$ are reproduced, as indicated. See text for details, and Fig.~\ref{Fig:param1} for comparison with the results obtained in the model-independent analysis.}}
\label{fig:epsilon_model}
\end{figure}

In the model-independent setup discussed in Sec.~\ref{sec:mi} and Sec.~\ref{sec:interplay}, the different annihilation rates and the effective CP-violation parameter were treated as independent free parameters. However, in a model in which such processes are realized, these parameters are often correlated, and are determined in terms of the common set of couplings and masses. Therefore, in order to understand whether the simple model described by Eq.~\ref{eq:lag} can accommodate the required values of the relevant physical parameters found in the previous section, we study in Fig.~\ref{fig:epsilon_model} the correlation between the effective CP-violation $\epsilon_{\rm eff}$, and the annihilation rates, as a function of the CP-violating phase $\theta$, and the relevant couplings $\lambda_2$ and $|\lambda|$. 

In Fig.~\ref{fig:epsilon_model}, we show contours of fixed effective CP-violation parameter $|\epsilon_{\rm eff}|$ (red solid lines), as a function of the complex phase $\theta=\arg(\lambda)$ and the effective pair-annihilation coupling $\lambda_2$. We have also shown the corresponding values of the annihilation rate $\langle \sigma v \rangle_A$ in both panels. The results are shown for two different DM mass values $m_\chi=1$ TeV (left panel) and $5$ GeV (right panel).  As we can see from this figure, the values of the annihilation rates and $\epsilon$ required to satisfy the DM relic abundance can be obtained in this model, as indicated by the dashed horizontal and vertical lines. This can be observed by comparison with the $\Omega h^2=0.12$ contour in Fig.~\ref{Fig:param1}, where the results were obtained in the model-independent analysis.

A few comments are in order. First of all, as mentioned earlier, in Fig.~\ref{fig:epsilon_model} we ensure $\mu / m_\chi << 1$ and $\mu_1 / m_\chi << 1$, for which our restriction to the class of $2 \rightarrow 2$ diagrams in Fig.~\ref{Fig:diagrams} remains valid.    Since the loop amplitudes in this model depend upon the coupling $\lambda_2$, the pair-annihilation process is necessarily present whenever the CP-violation in the semi-annihilation process is sufficiently large. Thus the first scenario with only the semi-annihilation process discussed in Sec.~\ref{sec:mi} is not obtained in this model, while the second scenario in Sec.~\ref{sec:interplay} with both semi- and pair-annihilations can be easily realized. Additional structures are therefore necessary to have loop graphs with sufficiently large imaginary parts, which do not induce significant tree-level pair annihilation~\cite{GGM2}.

\begin{figure}[htb!]
\begin{center}
\includegraphics[scale=0.55]{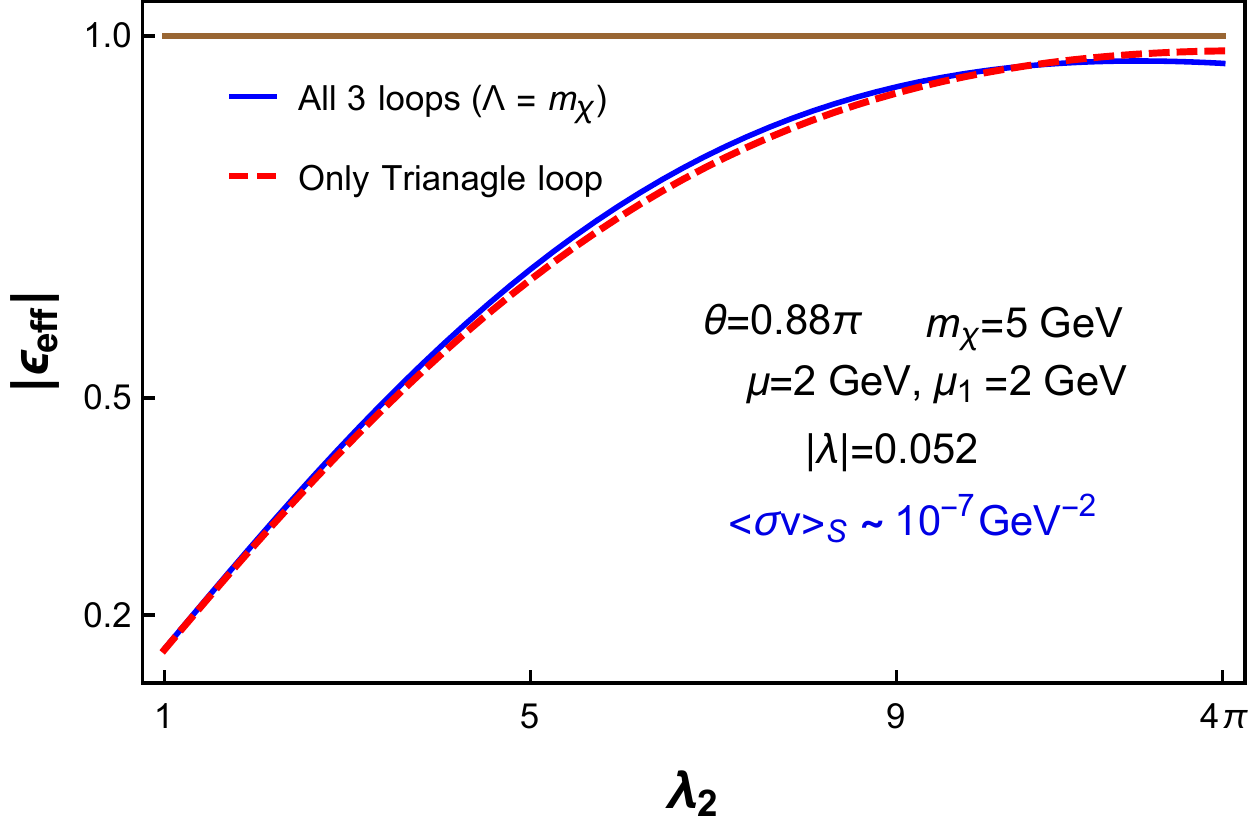} 
\caption{\small{\em The effective CP-violation parameter $|\epsilon_{\rm eff}|$, as a function of the pair-annihilation coupling $\lambda_2$. The results are shown for fixed values of all other parameters, including the DM mass value of $m_\chi=5$ GeV. The triangle loop gives the dominant contribution, as shown by comparison of $|\epsilon_{\rm eff}|$ computed only with the triangle loop (red dashed line), and all the three loops (blue solid line).}}
\label{Fig:eps_eff}
\end{center}
\end{figure}
We find that a scenario where the effective CP-violation parameter $|\epsilon_{\rm eff}|$ is close to unity, thereby leading to the present DM asymmetry $\eta \rightarrow 1$, can be realized in the model described by Eq.~\ref{eq:lag}, for values of the model parameters within perturbative limits. We show in Fig.~\ref{Fig:eps_eff} the variation of $|\epsilon_{\rm eff}|$ as a function of the coupling $\lambda_2$, which appears in all the relevant loop graphs in Fig.~\ref{Fig:diagrams}. The results are shown for fixed values of all the other parameters, including the DM mass value of $m_\chi=5$ GeV. Even though here the dimensionful parameters $\mu$ and $\mu_1$ are taken to be only a factor of two smaller than $m_\chi$, we have checked that only the $2 \rightarrow 2$ processes in Fig.~\ref{Fig:diagrams} dominate. We find that the triangle loop gives the dominant contribution, as shown by comparison of $|\epsilon_{\rm eff}|$ computed only with the triangle loop (red dashed line), and all the three loops (blue solid line). While the triangle loop ($L_1$ in Fig.~\ref{Fig:diagrams}) does not lead to any ultraviolet (UV) divergence, the other two loops (namely, $L_2$ and $L_3$) are divergent in the UV, and therefore $|\epsilon_{\rm eff}|$ computed with all three loops has a renormalization scale dependence, which is found to be rather weak. For our computations, we have set the renormalization scale to be $\Lambda = m_\chi$. We emphasize that even though in this example, the value of the pair-annihilation coupling $\lambda_2$ is large, the choice of parameters for which $|\epsilon_{\rm eff}| \rightarrow 1$ essentially belongs to the first scenario we studied, in which the semi-annihilation process almost entirely determines the present DM properties, including its asymmetry. This is because, for $|\epsilon_{\rm eff}| \rightarrow 1$, at the decoupling of the semi-annihilation process, the symmetric component is already negligible, and hence the subsequent pair-annihilation is largely irrelevant. Therefore, as is clear from Fig.~\ref{Fig:eps_eff}, we can achieve $|\epsilon_{\rm eff}|$ close to $1$ in this scenario, being within perturbative limit of all the relevant parameters including $\lambda_2$. Furthermore, we can also essentially realize, albeit in the presence of the coupling $\lambda_2$, the first scenario in which the semi-annihilation process almost entirely determines the present DM properties.

\section*{Acknowledgment}
We would like to thank Shigeki Matsumoto for many valuable discussions, a careful reading of the manuscript, and several important comments. We would also like to thank Ayres Freitas for a careful reading of the paper, and many important comments. SM would like to thank the Kavli IPMU, Japan, for hospitality, where part of this work was carried out. AG would like to thank the School of Physical Sciences, IACS, Kolkata for hospitality during the initial and final phases of this work. The work of AG is partially supported by the RECAPP, Harish-Chandra Research Institute, and the work of DG is supported by CSIR, Government of India, under the NET JRF fellowship scheme with award file No. 09/080(1071)/2018-EMR-I.

\end{document}